\documentclass[prb,aps,twocolumn,showpacs,floatfix,amsmath]{revtex4} 
\usepackage{epsfig,epsf,float}
     
\begin{document}

\title{Mean Field Theory of Collective Transport with Phase Slips} 
\author{Karl Saunders}

\altaffiliation{Present address:
  Centre for Condensed Matter Theory, Department of Physics, Indian 
  Institute of Science, Bangalore 560 012 India.}

\author{J.M. Schwarz}

\altaffiliation{Present address:
  Dept.\ of Chemistry and Biochemistry, 
  University of California, Los Angeles, CA 90095-1569.}

\author{M. Cristina Marchetti}
 
\author{A. Alan Middleton} 
\affiliation{Department of Physics, Syracuse University, Syracuse, NY 13244} 
 
\date{\today}  
\begin{abstract} 
The driven transport of plastic systems in various disordered backgrounds is
studied within mean field theory. Plasticity is modeled
using non-convex interparticle potentials that allow for phase slips.
This theory most naturally describes sliding charge density waves;
other applications include flow of colloidal particles or driven
magnetic flux vortices in disordered backgrounds. The phase diagrams
exhibit generic phases and phase boundaries, though the shapes of the
phase boundaries depend on the shape of the disorder potential.  The
phases are distinguished by their velocity and coherence: the moving
phase generically has finite coherence, while pinned states can be
coherent or incoherent.  The coherent and incoherent static phases can
coexist in parameter space, in contrast with previous results for
exactly sinusoidal pinning potentials.  Transitions between the
moving and static states can also be hysteretic. The depinning
transition from the static to sliding states can be determined
analytically, while the repinning transition from the moving to the
pinned phases is computed by direct simulation.
\end{abstract} 
\pacs{83.60.Bc,62.20.Fe} 
 
\maketitle
\section{Introduction} 
\label{Introduction}

The collective
dynamics of extended systems driven through quenched disorder is a
rich and challenging problem, with many experimental realizations.
Such systems include vortices in type II superconductors, charge
density waves in anisotropic conductors, domain walls in random
ferromagnets, and planar cracks in heterogeneous
materials.\cite{Fisher98} Much of the theoretical work to date has
focused on modeling these systems as extended {\em elastic} media.  In
these models the restoring forces are monotonically increasing
functions of the relative displacements, and the system is not allowed
to tear.  At zero temperature, overdamped elastic media subject to an
applied force $F$ and quenched disorder exhibit a nonequilibrium phase
transition from a pinned state to a sliding state at a critical value,
$F_T$, of the driving force.
\cite{DSF85}  The depinning transition, first fully studied for collective
models with disorder in the context of charge density waves, displays
the universal critical behavior of {\em continuous} equilibrium phase
transitions, with the mean velocity $v$ of the medium playing the role
of the order parameter.\cite{Fisher98,NF92}  For monotonic interactions, it
has been shown that the system's velocity is a unique function of the
driving force.\cite{nocross}  The sliding state is therefore unique
and there is no hysteresis or history dependence.  The depinning
transition of driven elastic media has been studied extensively, both
by functional renormalization group methods
\cite{NF92,Ertas96,Wiese02,Nattermann} and large scale numerical
simulations.\cite{LittlewoodCoppersmith86,LittlewoodCoppersmith87,middleton91,middleton93,MyersSethna93,RossoKrauthSimul}  Universality classes have
been identified, which are distinguished, for example, by the range of
the interactions or by the periodicity (or nonperiodicity) of the
pinning force. More recent work, while still focusing on elastic
media, has shown that the dynamics is quite rich well into the
uniformly sliding
state.\cite{koshelev94,balents95,giamarchi96,bmr98,moon96,pardo98}.

The elastic medium model is often inadequate to describe many real
systems which exhibit plasticity (due, for instance, to topological
defects in the medium) or inertial effects that violate the assumption
of overdamped equations of motion.  The dynamics of plastic systems
can be both spatially and temporally inhomogeneous, with coexisting
pinned and moving regions.\cite{coexistnote} The depinning transition
may become discontinuous (first order), possibly with macroscopic
hysteresis and ``switching'' between pinned and sliding
states. \cite{Maeda85,Maeda90, thorne03,foot_inertia} The theoretical
understanding of the dynamics of such ``plastic'' systems is much less
developed than that of driven elastic media.  A number of mean-field
models of driven extended systems with locally underdamped relaxation
or phase slips have been proposed in the literature,
\cite{Strogatz,levy92,levy94,NV97,Fisher98,MMP00,SF01,MCMKD02,SF03,MMSS03}
but many open questions remain.

Much of the original theoretical work on driven disordered systems
was motived by
charge density wave (CDW) transport in anisotropic
conductors, which display a nonlinear current-voltage
characteristic with a threshold voltage  for
collective charge transport. \cite{gruner,ThornePhysicsToday} 
It has been known for some time that 
the {\em elastic} depinning transition may not be
physically relevant to real CDW materials.
\cite{coppersmith90,SCAM91,ThornePhysicsToday} Coppersmith argued that in
elastic models with weak disorder,  
unbounded strains can
build up at the boundaries of  an atypically low pinning region,
resulting in large gradients of displacement that lead to the breakdown 
of the elastic model.\cite{coppersmith90}  Topological defects or phase slips will occur at
the boundaries of such a region, yielding a spatially nonuniform
time-averaged velocity. Theoretical and numerical studies of models that 
incorporate both phase 
and amplitude fluctuations of the CDW order parameter have indicated that
phase slips from large amplitude fluctuations can destroy the critical 
behavior. \cite{balents95,myers99,karttunen99} The
depinning may become discontinuous and hysteretic, or rounded, in the
infinite system limit.
Experiments show that varying the temperature of the CDW material
can lead to a transition from
continuous depinning to hysteretic depinning with sharp
``switching'' between pinned and sliding states.\cite{Maeda90,switch,Thorneexpt}
Furthermore, the observed correlation between the amplitude of
broadband noise and macroscopic velocity inhomogeneities also suggest
the presence of phase slips.\cite{broadband} 
It should be mentioned, however, that in many samples a substantial
amount of phase slips occurs
at the contacts,\cite{lemay98} while less clear evidence exists for
substantial
phase slip effects in the bulk. In general, CDW experiments display 
considerable sample-to-sample variability,\cite{thorne03}
making the comparison 
between theoretical
models and experiments quite challenging.

Related slip effects or plastic behavior have been proposed to explain
the complex dynamics of many other dissipative systems, including
vortex arrays in type-II superconductors. Simulations (mainly in two
dimensions) \cite{moon96,JensenBrassBerlinskySimul88,JensenBrassBrechetBerlinskySimul88,NoriSimul96,ReichardtOlsonSimul,GBD,FMM}, 
imaging \cite{pardo98,Marchevsky97,Troyanovski99,STM,lorenz}, and transport and noise 
experiments \cite{fingerprint,Hellerqvist96,maeda02} have shown  that
driven flux lattices often do not respond as elastic media.  Instead,
the driven lattice tears as small-scale topological defect structures
are generated and healed by the interplay of drive,
disorder and
interactions.
The tearing results in a ``plastic'' response, with highly defective
liquid-like regions flowing around the boundaries of pinned solid-like
regions.\cite{FMM}  This kind of response is most prominent in the
region near vortex lattice melting, where the so-called peak effect
occurs, i.e., the critical current shows a sudden increase with
temperature or applied field. Reproducible noise or ``fingerprint
phenomena'' have been observed in the current-dependent differential
resistance and attributed to the sequential depinning of various
chunks of the vortex lattice.\cite{fingerprint} Images of driven
vortex arrays in irradiated thin films of Niobium obtained by Lorentz
microscopy have shown clearly that vortex rivers flowing past each
other at the boundaries of pinned regions of the
lattice.\cite{lorenz} Scanning tunneling microscopy, which can
resolve individual vortices at high density, has also revealed a clear
evolution of the vortex dynamics with disorder strength.\cite{Troyanovski99} In samples
with weak disorder the vortex array was observed to creep coherently
along one of the principal crystal axes near the onset of motion. In
samples with strong disorder, the depinning is plastic and
the path of individual vortices can be followed as they meander
through the pinned crystal. Finally, as in the case of
CDWs, a correlation between plasticity and broadband noise has been
observed in several samples.\cite{maeda02}
Recently it has been argued that some of the observed behavior may be due 
to edge contamination effects that are responsible for the coexistence
of a metastable disordered phase and a stable ordered phase. \cite{paltiel,Paltiel02,marchevsky}
It is clear that more work is needed to understand the rich dynamics of these 
driven systems.

\begin{figure}
\begin{center}
\epsfxsize=8cm
\epsfbox{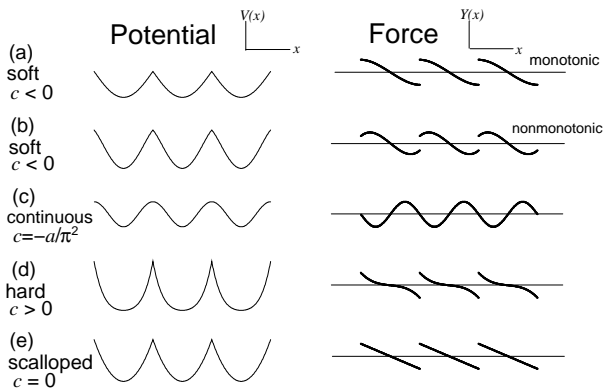}
\caption{ \label{potentials} Sketches 
of the pinning potentials and forces studied in this paper.  The
pinning forces are periodic with period $2\pi$ and the pinning
potential for a degree of freedom $\theta_i$ has minima at $\beta_i +
2n\pi$, for integer $n$. The cases are organized primarily by the sign
of $c$, with the pinning force $Y(x)=-ax-cx^3+O(x^5)$ for small
$x=\theta_i-\beta_i$. The coefficient of the harmonic part of the force
satisfies $a>0$.  The cases (a), (b) and (c) are for ``soft'' pinning
forces ($c < 0$); they differ near the potential maxima, corresponding to
monotonic, nonmonotonic, and continuous forces, respectively.  Case (d)
is a ``hard'' potential ($c>0$). The ``scalloped'' potential, case
(e), is precisely quadratic ($c=0$) in the interval $-\pi < x <
\pi$. The form of the potential especially affects the stability of
the coherently pinned phase and whether ``reentrant'' pinning is
possible upon increasing or decreasing the force.}
\end{center}
\end{figure}

In this paper we study the driven dynamics of a disordered medium with
phase slips, in order to better address questions about these and
related physical systems.  We restrict ourselves to systems which are
periodic along the direction of motion, such as CDWs, vortex lattices
or 2D colloids, and consider only the dynamics of a scalar
displacement field.  For concreteness, the model is described in the
context of driven CDWs, but it also applies to other driven systems
with pinning periodic in the displacement coordinate.  Assuming
overdamped dynamics and discretizing spatial coordinates, the dynamics
of the phase $\theta_i$ of each CDW domain is controlled by the
competing effects of the external driving force, the periodic pinning
from quenched disorder, and the interaction among neighboring domains.
Following the literature,\cite{Strogatz,Gorkov,OngMaki,inui88} phase
slips are introduced by modeling the interactions as a nonlinear sine
coupling in the phase difference of neighboring domains. The mean
field limit for this type of model has been studied by Strogatz,
Westervelt, Marcus and Mirollo\cite{Strogatz} for the case of the
smooth sinusoidal pinning force and was shown to exhibit a first order
depinning transition, with hysteresis and switching.  In this paper we
use a combination of analytical methods and numerical simulations to
obtain the nonequilibrium mean field phase diagram of the phase slip
model for a variety of pinning forces (see
Fig.~\ref{potentials}). Note that most of the pinning forces we
consider are discontinuous. This form of the force mimics the cusped
potentials that are the starting points for mean field theories that
best reproduce the finite-dimensional results.
The discontinuous pinning forces also reflect the
abrupt changes in the effective force (sum of elastic and pinning
forces) that occur when a neighboring region of the medium suddenly
moves forward. We find that
discontinuous forces, and even continuous nonsinusoidal pinning forces,
yield a rich nonequilibrium phase
diagram, with novel stable static phases that are not present for
exactly sinusoidal pinning forces.

In mean field theory, the nonequilibrium state of the system can be
described in terms of two order parameters. As the pinning potential
for each domain $i$ is periodic in
$\theta_i$, having minima at $\beta_i + 2\pi n$, for integer $n$, and taking 
the interactions to be periodic in the difference $\theta_i-\theta_j$
between neighboring phases with the same period,
a natural order parameter is the {\em
coherence} of the phases. This coherence
is measured by the amplitude $r$ of a
complex order parameter defined via
\begin{equation}
\label{coherence}
re^{i\psi}=\frac{1}{N}\sum_{j=1}^{N}e^{i\theta_j}\;,
\end{equation}
with $\psi$ a mean phase. 
In the absence of 
interactions among the phases or
external drive, the $\theta_i$'s are locked to the random phases,
$\theta_i=\beta_i$,
and the state is incoherent, with $r=0$.
In the opposite limit of very strong interactions  we
expect perfect coherence of the static state, with all phases becoming
equal and $r\to 1$ as the interactions become strong (or the pinning
becomes weak.)  Another order parameter is the
average velocity of the system, given by 
\begin{equation}
\label{meanvel}
v=\frac{1}{N}\sum_{j=1}^{N}\dot{\theta}_j(t)\;.
\end{equation}
The mean velocity is the order parameter for the transition between
static and moving phases.

\begin{figure}
\begin{center}
\epsfxsize=8cm
\epsfbox{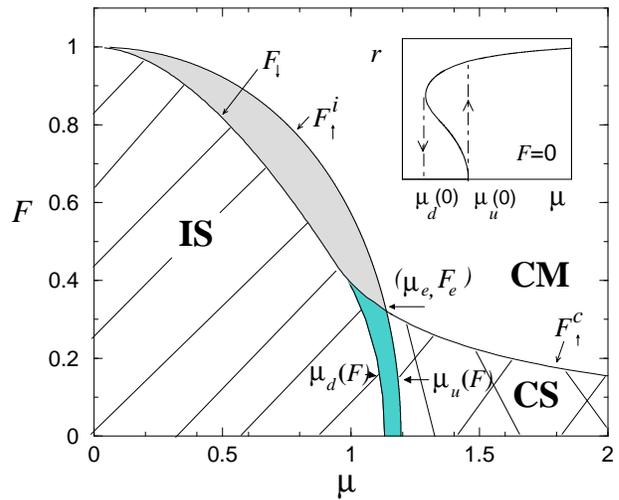}
\caption{\label{soft3}Phase diagram in the coupling-drive ($\mu$-$F$) 
plane for a discontinuous soft cubic pinning force of the type shown
in Fig.~\ref{potentials}b. The equation of motion is
Eq.~(\ref{mft_eom1}). The corresponding $Y(x)$ is given by
Eq.~(\ref{Y(x)}) with $a=15/(8\pi)$ and $c=-4a^3/27$. The strength of
the pinning is $h=1$ for all degrees of freedom. The diagonally lined
region indicates the IS phase, while the cross-hatched region
indicates the CS phase.  The light gray shaded region denotes the
region of coexistence of the CM and IS phases, while the medium gray
shaded region denotes the region of coexistence of the IS and CS
phases. The lines $F_{\uparrow}^i$ and $F_{\uparrow}^c$ are the forces
at which the system depins upon increasing the drive from the
incoherent and coherent static states respectively. The line
$F_{\downarrow}$ is the force at which a coherently moving system
stops upon lowering the drive. The point $(\mu_e,F_e)$ indicates where the 
static-moving transition goes from hysteretic to non-hysteretic. 
The curves $\mu_u(F)$ and $\mu_d(F)$
are the values of the coupling at which the static system makes the
transition to and from finite coherence states, respectively. The
inset displays the hysteresis in the coherence $r$ as the coupling
strength $\mu$ is varied at $F=0$. The transitions between the IS and
CS phases are first order in $r$. }
\end{center}
\end{figure}

The central results of this paper are the nonequilibrium phase
diagrams describing the static and moving phases, for the various
pinning forces shown in Fig.~\ref{potentials}.  The parameters for the
phase diagrams are the driving force $F$ and the strength $\mu$ of the
interaction between the domains. (For a phase diagram in the drive
force vs.\ pinning strength plane, see Sec.`\ref{Conclusions}.)
Although the precise shape of the phase boundaries depends on the
detailed form of the pinning potential, the types of phases and the
schematic topology of the phase diagram are general. This topology
and set of phases is exemplified in the phase diagram for the
discontinuous soft cubic pinning force (see Fig.~\ref{potentials}(b))
shown in Fig.~\ref{soft3}.  We find three distinct zero-temperature
nonequilibrium phases:
\begin{itemize}
\item an {\em incoherent static phase} (IS) at low drives and 
small coupling strengths,
with  $v=0$ and $r=0$;
\item a {\em  coherent static phase} (CS) at low drives and 
large coupling strengths,
with  $v=0$ and $r>0$;
\item a {\em coherent moving phase} (CM) at large drives,
with  $v>0$ and $r>0$.
\end{itemize}
We have investigated the possibility of an incoherent moving (IM) phase.  For 
continuous pinning forces, there is no IM phase.  For discontinuous 
pinning forces, we speculate that the IM phase is unstable {\em generically}.
(See Sec. V where the stability of a possible IM phase is discussed.)    

An important new feature of the phase diagram is the occurrence of a
{\em coherent static phase} at finite $F$. In contrast, for the
sinusoidal pinning force studied previously by Strogatz and collaborators 
\cite{Strogatz} the static state is always
incoherent (IS) for all finite values of the driving force and the CS
phase is only present at $F=0$.   

\begin{figure}
\begin{center}
\epsfxsize=8cm
\epsfbox{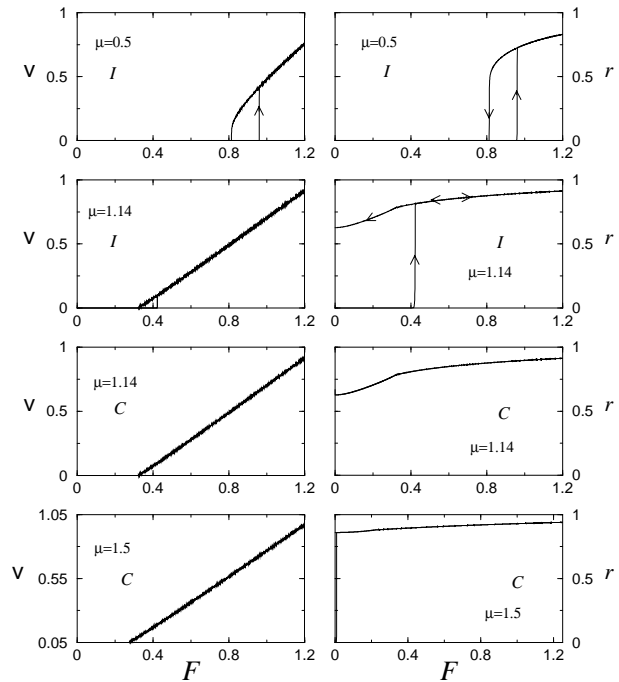}
\caption{\label{soft3_vfrf} Typical numerical results, found by integrating 
numerically the equations of motion (Eq.~(\ref{mft_eom1})), for the
behavior of the mean velocity $v$ and the coherence $r$ as the driving
force is slowly varied. For each pair of plots, the coupling $\mu$ is
held constant, while the drive force $F$ is raised from $F=0$ to
$F=1.2$ and then decreased.  The pinning potential is the same as for
Fig.~\ref{soft3}.  The top frames ($\mu = 0.5$) show
the hysteretic behavior between the IS and the CM phases, where the
coherence and velocity jump between zero and non-zero values at the
same locations. The next two sets of frames ($\mu=1.14$) are obtained
by preparing the system in the IS-CS coexistence region, starting from
either an initial incoherent ($I$) or coherent ($C$) state. When the
system is prepared in an incoherent state, the velocity and coherence
jump at at the same value of $F$ ($\approx 0.42$) as $F$ is raised,
but change continuously as $F$ is decreased, albeit with a change in
the slope $dr/dF$ at the repinning force $F\approx 0.32$, where $v$
goes to zero.  When the system is prepared in a coherent state, there
is no hysteresis and $v$ and $r$ are continuous, though $r$ again
shows a singularity at depinning.  The bottom frames ($\mu=1.5$)
display the behavior at the continuous depinning transition from the
CS phase. The results are similar to those for $\mu=1.14$, when
starting from the coherent state ($C$). In general, depinning from the
coherent state is continuous and non-hysteretic, while depinning from
the incoherent state is discontinuous and hysteretic. Numerical
evidence for the hysteresis does not change over the size ranges
studied, strongly suggesting that these simulations
accurately represent the infinite-volume
limit. }
\end{center}
\end{figure}

The location of the transitions between these phases depends on the
system's history.
Changing the coupling $\mu$ at fixed drive $F$ can give a
hysteretic transition between incoherent and coherent static
phases, as shown in the inset of Fig.~\ref{soft3} for
$F=0$.  Fig.~\ref{soft3_vfrf} shows
the behavior of both the mean velocity and the coherence
as $F$ is first increased and then decreased across
the boundaries between static and moving phases of Fig.~\ref{soft3}, while
keeping $\mu$ fixed.
The most important features of the phase diagrams are:
\begin{itemize}
\item
{\em The transition between the IS and CS phases is generally
discontinuous.} The region of coexistence of coherent and 
incoherent static states is bounded by curves  $\mu_d(F)$ and $\mu_u(F)$ 
(or equivalently $F_d(\mu)$ and $F_u(\mu)$.)
When the coupling strength $\mu$ is increased at fixed
$F$ within the static region, the system jumps from an incoherent to a
coherent state at the critical value $\mu_u(F)$, with a discontinuous
change in $r$ (see inset of Fig.~\ref{soft3}).  When $\mu$ is ramped
back down, the coherent static state remains stable down to the lower
value $\mu_d(F)$. The boundaries $\mu_d(F)$ and $\mu_u(F)$ 
coincide for the piecewise linear pinning force. 
In this case the transition is still
discontinuous, but not hysteretic.  An exception to this general
behavior is found for the hard pinning potential at very small values
of $F$, where the transition between coherent and incoherent static
states is continuous.
\item{\em The depinning to the moving phase is discontinuous and
hysteretic when the system depins from the IS phase (except when $\mu=0$).}
When $F$ is increased adiabatically from zero at fixed $\mu$ for 
a system prepared in the IS phase, both the velocity and the coherence 
jump discontinuously from zero to a finite value at $F_\uparrow^i(\mu)$.
For an example, see the top frames of Fig.~\ref{soft3_vfrf}.
When the force is ramped back down from the sliding state the
system gets stuck again at the lower value $F_\downarrow(\mu)$. 
\item{\em The depinning to the moving phase is generally continuous when
the system depins from the CS phase.}  In this case both the velocity
and the coherence change continuously at the transition, although they
may be non-analytic functions of the control parameters. An example of
this behavior is displayed in the bottom frames of
Fig.~\ref{soft3_vfrf}.  An exception is found for piecewise
linear pinning forces (case (e) of Fig.~\ref{potentials}) for
$\mu\agt\mu_u$.
\item {\em  For continuous pinning forces, the depinning threshold
$F_\uparrow^c(\mu)$ vanishes for $\mu$ above a critical $\mu_T$.}
In contrast, discontinuous pinning forces exhibit a finite depinning 
threshold for all finite values of $\mu$ with $F_\uparrow^c(\mu)$
decreasing with increasing $\mu$.
\end{itemize}
Analytical expressions have been obtained for the critical lines 
$F^c_\uparrow(\mu)$ and
$F^i_\uparrow(\mu)$, which give the depinning force values for the
coherent and the incoherent static phases, respectively, as well as for the phase
boundaries $\mu_d(F)$ and $\mu_u(F)$, which separate the coherent and
incoherent static phases. Numerical simulations of finite mean-field
systems have also been used to obtain these boundaries, confirming the
analytic stability criteria.  The repinning curves
($F_\downarrow(\mu)$), where moving solutions stop upon lowering the
drive $F$, have been determined numerically.

Part of the motivation for our work comes from the well-known result
that the mean field critical exponents for the depinning transition in
purely elastic models depend on the details of the pinning force. For
instance, the exponent $\beta$ controlling the vanishing of the mean
velocity $v$ with driving force at threshold, $v\sim(F-F_T)^\beta$,
has a mean field value $\beta=3/2$ for generic smooth continuous
pinning forces and $\beta=1$ for a discontinuous piecewise linear
pinning force (Fig.~1(e)).\cite{AAMthesis} Using a functional RG (FRG)
expansion in $4-\epsilon$ dimensions, Narayan and Fisher showed
\cite{NF92} that the discontinuous force captures a crucial intrinsic
discontinuity of the large scale, low-frequency dynamics. The FRG
calculations give $\beta=1-\epsilon/6+{\cal O}(\epsilon^2)$, in good
agreement with numerical studies in two and three dimensions.  The
mean field elastic medium also has zero depinning field, $F_T=0$, for
small pinning strengths $h$, in contrast with finite dimensional
simulations and predictions for a finite depinning field in any
dimension based on Imry-Ma/Larkin-Ovchinnikov and rare region
arguments.\cite{DSF85} The RG calculation and the numerics show that a
discontinuous pinning force must be used in the mean field theory to
incorporate the inherent jerkiness of the motion of finite-dimensional
systems at slow velocities.  Although there is no reason to believe a
priori that the same will hold for models with phase slips, it is
clearly important to understand how the properties of the pinning
potential affect the nonequilibrium phase diagram of the
model. Furthermore, for large coupling strength $\mu$ and bounded
pinning force the phase slip model reduces to the elastic model, where
the nature of the pinning force strongly affects the mean field
theory.

For further applications and connections, we note that models of
driven disordered systems with nonmonotonic interactions are also
relevant for arrays of nonlinearly coupled oscillators.  An example is
the Kuramoto model used to describe the onset of synchronization in
many biological and chemical systems.\cite{kuramoto} The model
consists of a large number of oscillators with random natural
frequencies and a sinusoidal coupling in their local phase
differences.  Although there is no external drive, this model can
exhibit a transition to a synchronized phase as the strength of the
coupling is increased. In this phase, all the degrees of freedom
oscillate at a common frequency.  In the Kuramoto model the natural
frequency acts as a random driving force that varies for each
oscillator, but there is no random pinning. The model considered here,
in contrast, consists of coupled phases, or oscillators, in a random
pinning environment at fixed (constant) drive.  The onset of coherence
(either in a moving or in a static state) corresponds to the onset of
the synchronization in the Kuramoto model.

We conclude this introduction by briefly summarizing the remainder of
the paper.  In Sec.~\ref{Model} we describe the model of driven CDWs
with phase slips and introduce the mean field limit. In
Sec.~\ref{Static States} we obtain the static solutions of the mean
field model at $F=0$ for the selection of pinning forces shown in
Fig.~\ref{potentials}.  We show that the existence of a transition
between incoherent and coherent static states can be inferred
perturbatively. A full non-perturbative treatment is then applied to
understand the nature of the transition.  In Sec.~\ref{IS phase for F
nonzero} we consider static states at finite drive. Again, the region
of stability of the incoherent static phase can be established by
perturbation theory, but the nonperturbative treatment described in
Sec.~\ref{nonzero F COV} is needed to map out all the static states
and their boundaries of stability to the moving state. The resulting
phase diagrams for the various classes of pinning forces are discussed
in Sec.~\ref{nonzero F COV}; the analytic calculations supporting
these phase diagrams are presented in Appendices \ref{calcA} and
\ref{calcB}.  As the analytic treatment we present here is restricted
to finding boundaries starting from the static phases, the lower
boundary $F_\downarrow(\mu)$ of the hysteretic region where static and
moving state coexist has been obtained numerically.
Sec.~\ref{Averaging over disorder} addresses the effect of a broad
distribution of pinning strengths.  We conclude in
Sec.~\ref{Conclusions} with a discussion of the results and avenues
for further studies.

\section{The model}
\label{Model}
Though the results of our analysis are more general, we motivate
the model with a detailed discussion of the physics of CDWs. The
general ideas of phase slip also apply to other systems, most directly
to coupled layers of vortices, where the vortices are confined to the
planar layers, or to colloidal
particles in a disordered background.

A CDW is a coupled periodic modulation of the electronic density and
lattice ion positions that exists in certain quasi-one dimensional
conductors, due to an instability of the Fermi surface.
The undistorted CDW state is a periodic condensate
of electrons, characterized by a complex order parameter, with an
amplitude $\rho_1$ and a phase $\theta$. 
The electron density can be expanded as
$\rho_{e}(x)=\rho_0+\rho_1\cos[Q_cx+\theta(x)]$, with $Q_c=2k_F$,
$k_F$ being the Fermi wave vector.   The phase $\theta(x)$ describes the
position of the CDW with respect to the lattice ions and is a constant for an
undistorted CDW.   When $Q_c$ is
incommensurate with the lattice, the CDW can ``slide'' and CDW
transport can be modeled using uniform translations and small
gradients of $\theta(x)$, to a first approximation.  An applied
electric field exceeding a threshold field causes the CDW to slide
relative to the lattice at a rate $\partial_t\theta$, giving rise to a
CDW current.  Amplitude fluctuations (changes in $\rho_1$) are often
neglected because they cost a finite energy, while a vanishingly small
energy is required to generate long-wavelength phase excitations, in
an ideal crystal.  This has led to the well known phase-only model of
CDW dynamics introduced by Fukuyama, Lee and Rice (FLR) that
incorporates long wavelength elastic distortions of the
phase.\cite{FLR} Strong disorder or regions of unusually low pinning
can lead to large strains, however, so that the amplitude can no
longer be regarded as constant. Large local strains can be relieved by
a transient collapse of the CDW amplitude.  One approach to describe
such a strongly strained system is a ``soft spin'' model that
considers the coupled dynamics of both phase {\em and} amplitude
fluctuations. This has been attempted by some
authors,\cite{balents95,myers99,karttunen99} but generally leads to
models that have to be treated numerically.  An alternative, more
tractable approach, is to continue to treat the amplitude as constant,
while modifying the interaction between phases. This modification
should incorporate the crucial feature that the phase becomes
undefined at the location where the amplitude collapses. At a strong
pinning center, phase distortions can be large and lead to the
accumulation of a large polarization that suppresses the CDW
amplitude, driving it toward collapse. When the distortion is released
through an amplitude collapse, the phase abruptly advances of order $\approx
2\pi$, while the amplitude quickly regenerates.\cite{inui88} This
process is known as phase slippage in superconductors and superfluids,
although it is modified in CDWs because of the physical coupling to
the phase.  On time scales large compared to those of the microscopic
dynamics, it can be described approximately as a ``phase slip'': an
instantaneous $2\pi$ (modulo $2\pi$) hop of the CDW phase.  Following
the literature, phase slips will be modeled here as phase couplings
periodic in the phase difference between neighboring domains. This
leads to a simple model that can be analyzed in some detail.

When modeling CDWs, especially numerically, displacements and
amplitudes are coarse-grained to a length scale of order the
Imry-Ma-Larkin-Ovchinikov length. At and below this scale, the CDW
behaves roughly as a rigid object, referred to as a correlated domain.
This domain is taken to move uniformly and is acted upon by
driving forces and interactions with neighboring domains and the pinning
potential.  The continuum space description is replaced with a
discrete set of degrees of freedom. The coarse-grained equation of
motion for the phase $\theta_i$ of a CDW domain $i$ is given by
 \begin{equation}  
{\dot \theta_i}= F + \mu\sum_{\langle j \rangle} \sin (\theta_j-\theta_i) 
+ h_i Y(\theta_i-\beta_i)\;, 
\label{eom} 
\end{equation} 
where the dot denotes the time derivative (we have chosen to scale
time so that the damping constant is unity) and $F$ is the driving
force.  The second term on the right hand side of Eq.~(\ref{eom})
represents the force due to the coupling to other domains, where
$\langle j\rangle$ ranges over sites $j$ that are nearest neighbor to
$i$ and $\mu$ is the coupling strength. The third term is the pinning
force which tends to pin the phase of each domain to a random value
$\beta_i$ uniformly distributed in $[-\pi, \pi]$.  The function $Y(x)$
is periodic with period $2\pi$ and represents the pinning forces. We
choose $Y(0)=0$ to fix the location of the minimum of the pinning
potential and set $Y''(0)=0$ to maintain reflection symmetry in the
absence of an external drive.  As the potential is minimized
at $\theta_i=\beta_i$, $Y'(0)<0$. The random pinning strengths $h_i$ are
independently chosen from a probability distribution $\rho(h)$.

The key difference between our model equation of motion and the well
known FLR elastic model of driven CDWs is in the form of the coupling
between domains. Instead of assuming a linear elastic force
$\sim\sum_{<j>}(\theta_j-\theta_i)$ between neighboring domains, we
have assumed a non-linear, sinusoidal coupling that allows for phase
slip processes.  For large phase distortions (exceeding $\pi$) the
restoring force in Eq.~(\ref{eom}) becomes negative and the phases
slip by an amount $2\pi$ relative to one another in order to relax the
strain.

The starting point for many finite-dimensional theories is the 
mean field picture where every local phase (or domain) is equally coupled to 
every other.  In this limit, the equation of motion (\ref{eom}) becomes 
\begin{equation}  
{\dot\theta}(\beta, h)=F - u \sin (\theta-\psi) + h Y(\theta-\beta)\;,
\label{mft_eom1} 
\end{equation} 
where 
\begin{equation}  
u\equiv\mu r\;,
\label{u_definition} 
\end{equation} 
measures the effective strength of coupling between the
domains and the mean field, with $r$ and $\psi$ defined in 
Eq.~(\ref{coherence}). This coupling will only be non-zero if
there is some coherence between the phases of different domains, i.e.,
if $r \neq 0$. For simplicity, we have dropped the
subscripts, labeling each phase by the values of
$\beta$ and $h$, which are now both continuous
variables. The $\beta$ are distributed uniformly in $[-\pi,\pi]$ and 
the $h$ have the distribution $\rho(h)$.

The self-consistency condition for the mean field theory is given by
\begin{equation} 
r e^{i\psi}={1\over2\pi}\int_{-\pi}^{\pi}d \beta \int dh\, \rho(h) e^{i\theta(\beta,h)}\;.
\label{self_consistency1} 
\end{equation} 
In this paper we will for the most part consider a narrow distribution
of pinning strengths, i.e., $\rho(h)=\delta(h-1)$. The effects of a
broad distribution $\rho(h)$ will be addressed in Sec.~\ref{Averaging
over disorder}.

When the phases are not coupled ($\mu=0$), the equation of motion
reduces to that of a single particle, which depins at the single
particle threshold force, $F_{\rm sp}$, given by the maximum pinning
force. Note that when the coherence $r$ is zero, then $u=0$, and the
system may also depin at $F_{\rm sp}$ for a finite value of $\mu$, as long
as $r$ remains zero.

\section{Static States For Zero Drive}
\label{Static States}

We first consider static solutions (${\dot\theta}=0$) to
Eq.~(\ref{mft_eom1}) for the case of zero drive ($F=0$).  These
solutions are the first step in determining the phase diagram and
their derivation introduces most of the techniques and concepts used
for non-zero drive. When $F=0$, the coherence $r$ is determined by
competition between two effects: the disordering effect of the
random impurities and the ordering tendency arising from the coupling
of each degree of freedom to the mean field. The outcome of this
competition gives the $\mu$-dependence of $r$.  At zero drive, the
system can exist in one of two possible phases: the disordered ($r=0$)
IS phase and the ordered ($r>0$) CS phase. These phases can coexist.
In this section we examine the nature of the transition between these
two phases obtained by varying $\mu$ at $F=0$. We find that the nature
of the transition depends on the shape of the pinning force, $Y(x)$.

For static solutions at zero drive, the equation of motion (\ref{mft_eom1}) 
reduces to the condition that the pinning force on each degree of freedom 
be balanced by the force due to deformations from coupling to the mean field,
\begin{equation}  
0=- u \sin (\theta-\psi) + h Y(\theta-\beta)\;,
\label{mft_static} 
\end{equation} 
where the reader is reminded that the effective coupling $u$ results
from the coupling strength $\mu$ and coherence $r$, $u=\mu r$.  For
any value of $\mu$ this equation has the trivial solution
$\theta=\beta$, $r=u=0$, where all phases rest at the minima of their
pinning potentials and the coherence and effective coupling are both
zero. It turns out, however, that such a static incoherent solution
becomes unstable above a characteristic value of the coupling strength
$\mu$.

In order to study the competition between the impurity disordering and
mean-field ordering effects, it is useful to rewrite the equation in
terms of the deviation $\delta$ of each phase from its value in the
disorder dominated incoherent state, $\delta\equiv\theta-\beta$.
A direct and important symmetry of the solution of Eq.~(\ref{mft_static})
is global phase invariance, which holds due to the uniform choice of $\beta$.
In the static state, this statistical rotational invariance means that we can
simply fix $\psi$ to be zero.  Given a solution with $\psi = 0$,
all related solutions with $\psi\not=0$ can then be
obtained by letting $\theta\rightarrow\theta+\psi$ and
$\beta\rightarrow\beta-\psi$. With this transformations, and
specializing to the case of fixed pinning strength, $h=1$, the force
balance equation becomes
\begin{equation} 
0= - u \sin (\delta+\beta) +  Y(\delta)\;. 
\label{force_balance1} 
\end{equation} 
To solve this force balance equation, we need to determine $u$
self-consistently.  The self-consistency condition
Eq.~(\ref{self_consistency1}) can be rewritten, by separating out its real
and imaginary parts, as
\begin{equation} 
r ={1\over2\pi}\int_{2\pi}d \beta \cos(\delta+\beta)\equiv f(u)\;,
\label{self_consistency1a} 
\end{equation} 
where we have implicitly used Eq.~(\ref{force_balance1}) to solve for $\delta$
as a (possibly multi-valued) function of $\beta$ and $u$ to define a
function $f(u)$ as the above average over $\beta$,
and 
\begin{equation} 
0 =\int_{2\pi}d \beta \sin(\delta+\beta)\;.
\label{self_consistency1b} 
\end{equation} 
Next, we will use a straightforward linear stability analysis to show
that the IS ($r=0$) phase becomes unstable to the CS ($r> 0$) phase
above a critical value $\mu_u$ of the coupling strength. A
perturbative calculation of $r(\mu)$ allows us to establish that this
transition from the IS to the CS phase is continuous or hysteretic,
depending on the shape of the pinning potential near its
minimum. We will then obtain the full solution $r(\mu,F=0)$ for a
variety of pinning forces.

\subsection{Stability of the Incoherent Static Phase}
\label{Linear Stability}

To investigate the linear stability of the IS phase, we calculate the
time evolution of a configuration near the static
solution $\delta(\beta)=0$. 
A convenient perturbed configuration is
$\delta(\beta,t=0)=-\epsilon(0)\sin\beta$
with $\epsilon(0)\ll 1$.
This perturbation gives non-zero
coherence while maintaining $\psi = 0$ and reflects
the most rapidly growing eigenvector in the stability analysis, with
$\delta(\beta,t)=-\epsilon(t)\sin\beta$ to lowest order in $\epsilon$.
By Eq.~(\ref{self_consistency1a}), the coherence of the perturbed state is
\begin{eqnarray} 
r&=& {1\over2\pi}\int_{-\pi}^{\pi}d \beta
\cos(\beta-\epsilon\sin\beta) \;,\nonumber\\ &=&\epsilon/2+O(\epsilon^2)\;.
\label{perturbed_r} 
\end{eqnarray} 
The equations of motion Eq.~(\ref{mft_eom1})
then give
\begin{eqnarray}  
{\dot\delta}&=&-\mu r \sin(\beta+\delta) + hY(\delta)\nonumber\\
&=&-\mu(\epsilon/2)\sin\beta+hY'(0)\delta+O(\epsilon^2)\nonumber\\
&=&\left(\frac{\mu}{2}+hY'(0)\right)\delta+O(\epsilon^2)\;.
\label{pertubed_eom} 
\end{eqnarray} 
As 
$r$ and $\delta$ are both proportional
to $\epsilon$ (to lowest order), it immediately follows that
${\dot r}\approx \left({\mu\over2}+hY'(0)\right)r$.
The critical value of $\mu$ for linear stability is therefore
\begin{equation}
\mu_u = -2 h Y'(0)\;.
\end{equation}
For coupling strength $\mu > \mu_u$,
the perturbed coherence grows
and the IS phase is linearly unstable to a CS phase.
At larger $\mu$, the interactions
that drive the $\theta$ towards
a coherent state are larger in magnitude than
the restoring force for the individual $\theta$.
Note that $\mu_u$ depends only on the strength of the
pinning force at the minimum of the pinning potential.

\subsection{Perturbation Theory}
\label{Perturbation Theory}

The onset of coherence for $\mu$ just above $\mu_u$ can be studied
perturbatively by assuming that both the phase $\delta$ and the
coherence $r$ are small in this region. Near $\delta=0$ the pinning
force can quite generally be written as a power series in $\delta$,
\begin{equation} 
Y(\delta)= -a \delta- b \delta^2 -c\delta^3 +... \;,
\label{general_pinning1} 
\end{equation} 
with $a=-Y^\prime(0)>0$. For small $r$, and hence $u$,
one can expand $\delta(\beta,u)$ in powers of
$u$,
\begin{equation} 
\label{delta_exp}
\delta(\beta,u)=u\delta_1(\beta)+u^2\delta_2(\beta)+u^3\delta_3(\beta)+...  \;.
\end{equation} 
Substituting these terms into the force
balance equation Eq.~(\ref{force_balance1}), and equating terms of
the same order in $u$, we obtain
\begin{subequations}
\begin{eqnarray}  
\delta_1(\beta)&=&-{\sin \beta \over a}\;,\label{expanded_delta_2a}\\ 
\delta_2(\beta)&=&{\sin \beta \cos(\beta)\over a^2} - {b \sin^2\beta
\over a^3}\;,\label{expanded_delta_2b}\\
\delta_3(\beta)&=&\left({c\over a^4}+{2 b^2 \over a^5} - {1\over2a^3}\right)
\sin^3\beta\nonumber\\ &-& {\sin\beta\cos^2\beta\over a^3} +
{3b\sin^2\beta\cos\beta\over a^4}\;.\label{expanded_delta_2c}
\label{expanded_delta_2}
\end{eqnarray}
\end{subequations}
Substituting the expanded $\delta(\beta,u)$ into
Eq.~(\ref{self_consistency1a}) and evaluating the integrals to each
order in $u$ we find
\begin{equation}
f(u)=u r_1 + u^2 r_2 + u^3 r_3 + \dots \;,
\end{equation}
with
\begin{subequations}
\begin{eqnarray}  
r_1&=&{1\over2a}\;,\label{f(u)_2a}\\ 
r_2&=&0\;,\label{f(u)_2b}\\
r_3&=&-{3\over8}\left({ac-2b^2\over a^5}\right)\;.\label{f(u)_2c}
\label{expanded_f(u)_2}
\end{eqnarray}
\end{subequations}
Finally, the coherence $r$ is given by the solution of
\begin{equation} 
r=f(r\mu)=(r\mu)r_1+(r\mu)^3r_3+\dots \;.
\label{expanded_f(u)} 
\end{equation} 
For simplicity of discussion we specialize to pinning potentials with
reflection symmetry and choose $b=0$ (although the non-zero $b$ result
will prove useful in the analogous finite $F$ perturbation
theory). Then $r_3=-3c/(8a^4)$ and the nonvanishing
solution for the coherence can be written as
\begin{equation}
r(\mu)
=\begin{cases}
\left(\frac{\mu_u^4}{3|c|\mu^3}\right)^{1/2}\left({\mu_u-\mu\over\mu_u}\right)^{1/2} &
c<0,\\
\left(\frac{\mu_u^4}{3c\mu^3}\right)^{1/2}\left({\mu-\mu_u\over\mu_u}\right)^{1/2} & 
c> 0,
\end{cases}
\label{small_r(mu)2}
\end{equation}
where $\mu_u=2a$.

\begin{figure} 
\begin{center}
\epsfxsize=7cm
\epsfbox{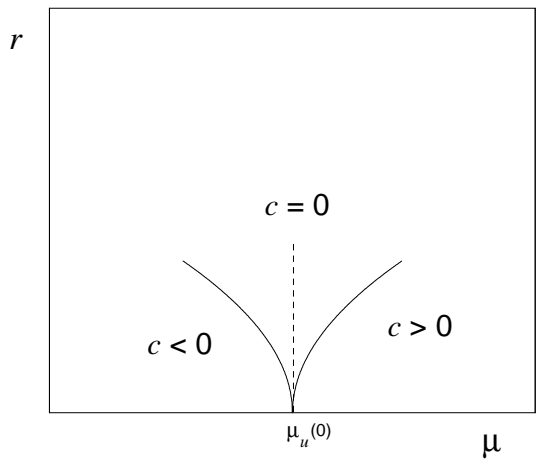}
\caption{The behavior of the coherence $r$ for couplings
$\mu\approx \mu_u$, i.e., near the instability point of the incoherent
static phase (IS) at $F=0$. The three curves show $r$ with pinning force
$Y(\delta\ll1)=-a\delta-c\delta^3$ for $c$ positive (hard pinning
potential), negative (soft pinning potential) and zero (piecewise linear
pinning force.)
}
\label{r(mu-mu_c)}  
\end{center}
\end{figure}

The behavior of $r(\mu)$ for $\mu\approx \mu_u$ and the nature of the
transition between the IS and CS phases are controlled by the sign of
the coefficient $c$ of the cubic term of $Y(\delta)$. The three types
of behavior that can occur are shown in Fig.~\ref{r(mu-mu_c)}.  For
$c>0$, corresponding to a ``hard'' pinning potential that grows more
steeply than a parabola near its minimum, the coherence $r$ grows
monotonically with increasing $\mu$, with
$r\sim(\mu-\mu_u)^{1/2}$. This indicates a continuous transition
at $\mu=\mu_u$ between the IS and CS phases. On the other hand, when
$c<0$, corresponding to a ``soft'' pinning potential, the coherence
starts out with a negative slope at $\mu_u$ and grows with {\em
decreasing} $\mu$.  We expect this solution to be unstable, indicating
that the transition from the IS phase to the CS phase occurs with a
discontinuous jump in $r$ from $r=0$ for $\mu<\mu_u$ to a non-zero
value of $r$ for $\mu>\mu_u$ on a stable upper branch not accessible
in perturbation theory. In fact we show below that when $\mu$ is
decreased back down through $\mu_u$ from the CS phase $r$ will remain
non-zero down to a lower value $\mu_d < \mu_u$, indicating a
hysteretic transition between the IS and CS phases.
In the marginal case of piecewise linear pinning forces with $c=0$,
i.e., $Y(\delta)= -a\delta$ near $\delta=0$, there is a discontinuous
jump $r(\mu)$ at $\mu=\mu_u$. In this case the perturbation theory
breaks down and the solution must be obtained by the method described
in Sec.~\ref{COV}. This calculation will show that no hysteresis
occurs in the case of strictly linear pinning force. We stress that
the transition from the IS to the CS state at $F=0$ is controlled
entirely by the shape of the pinning potential {\em near its
minimum}. Specifically, the behavior is unaffected by the existence of
discontinuities in the pinning force at the edges of each pinning
well.

\subsection{Beyond Perturbation Theory:
The General Static ${\bf r(\mu,F=0)}$ Solution}
\label{COV}

In this section we outline a non-perturbative method for calculating
the integral $f(u)$ used in the self-consistency
equation, Eq.~(\ref{self_consistency1a}). This allows for
the determination of the coherence $r$ for all values of $\mu$.  In
addition to confirming the perturbative results obtained above, this
method allows the precise study of the discontinuous and hysteretic
transitions between the IS and CS phase, which cannot be done within
perturbation theory.

To obtain $f(u)$ by direct integration over $\beta$ in
Eq.~(\ref{self_consistency1a}) one would need to solve the transcendental
equation, Eq.~(\ref{force_balance1}), for $\delta(\beta,u)$. Such
a solution cannot in general be obtained analytically. Hence we take an
alternative approach in which we solve Eq.~(\ref{force_balance1}) for
$\beta(\delta,u)$ and integrate over $\delta$, rather than $\beta$,
i.e.,
\begin{eqnarray}
r={1\over2\pi}\int d \delta\left({\partial \beta \over \partial \delta}\right) \cos(\delta+\beta(\delta,u))\;.
\label{r(u)_1}
\end{eqnarray}
The change of variable in Eq.~(\ref{r(u)_1}) provides an important
simplification that allows us to calculate analytically the coherence
of the undriven static state for a general pinning potential.
This simplification does rely on understanding the subtleties of how
$\delta$ depends on $\beta$, as $\delta$ can be multivalued function of
$\beta$. The history of the sample can determine which branch(es) are
included in the configuration.

For a given $u$,
there is an infinite set of  solutions to  Eq.~(\ref{force_balance1}).
We index each with an integer $n$:
\begin{eqnarray}
\beta_n(\delta,u)= -\delta + n \pi + (-1)^n\sin^{-1}(Y(\delta)/u) \;,
\label{delta(beta)_maintext}
\end{eqnarray}
where we choose the $[-\pi/2,\pi/2]$ branch for $\sin^{-1}(x)$.
The range for $\delta$ is constrained to $-\delta_{\max}(u)\leq \delta
\leq \delta_{\max}(u)$, with $\delta_{\max}(u)\equiv Y^{-1}(u)$.

\begin{figure}
\begin{center}
\epsfxsize=8cm
\epsfbox{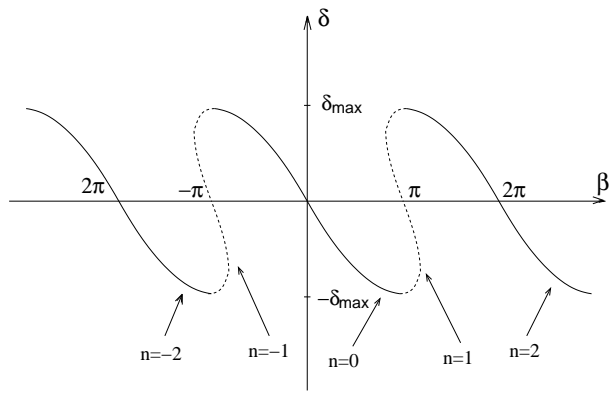}
\caption{
A sample plot of $\delta$, the displacement of a degree
of freedom from the minimum of the pinning potential, versus the
pinning phase $\beta_n(\delta)$ for branch numbers $-2\leq n \leq
2$. The
solid portions correspond to even $n$, while the dashed portions
correspond to odd $n$.
The global phase $\psi$ is chosen to be zero. Here, the
effective interaction is large enough, $u>a$, that $\delta(\beta)$ is
multivalued. The maximum magnitude of $\delta$ is denoted by $\delta_{\max}$.  
}
\label{beta)delta_fig}
\end{center}
\end{figure}

The calculation of the average in Eq.~(\ref{r(u)_1}) is easily
carried out when $u<a$, where the phase is single valued.  For values of $u
>a$ the function $\delta(\beta)$ is multivalued, allowing for the
existence of many metastable static configurations at fixed $u$.
Figure (\ref{beta)delta_fig}) shows one such multivalued
$\delta(\beta)$. Because of the metastability, 
the coherence can vary over some range.  For a fixed $u$, the range in 
coherence results in a range of couplings $\mu$.  
When $u\geq a$ and $\delta(\beta)$ is multivalued, one chooses
the (stable) branch of the $\beta(\delta)$ curve that is consistent
with the particular metastable state one wishes to describe and also
ensures that $\psi=0$, or equivalently that
Eq.~(\ref{self_consistency1b}) is satisfied.  For simplicity and
correspondence with ``typical'' sample preparation, we focus on those
metastable states accessed by adiabatically increasing $u$ from
zero.\cite{adiabatic_u_footnote} These correspond, for a given $u$, to
the solid portions of the curve shown in Fig.~(\ref{beta)delta_fig}).
The details of the calculation for the scenario of adiabatically
increasing $u$, which selects one branch, are given in Appendix A.  It
is relatively straightforward to show that for a given $u$ these are
the states which have the largest coherence. This selection of
largest-$u$ states is
consistent with our numerical calculations. Note that the form of the
$\delta(\beta)$ curve and the discussion of multiple solutions is
formally quite similar to parts of the calculation for the purely
elastic case, though the physical motivation is rather
different.\cite{DSF85}

\begin{figure}
\begin{center}
\epsfxsize=8cm
\epsfbox{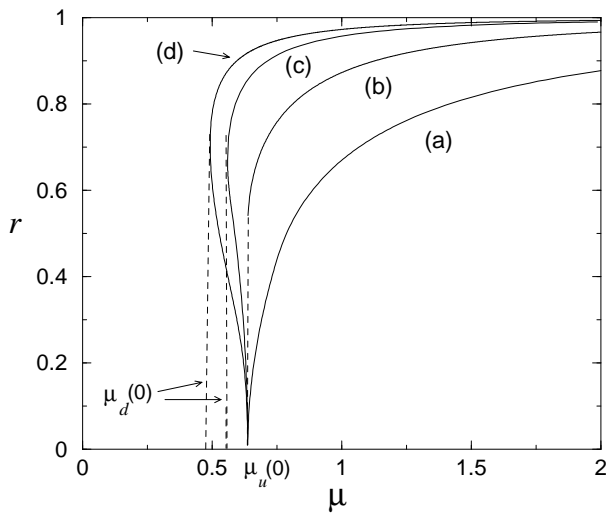}
\caption{The coherence of the static state at $F=0$ as a function of the
coupling strength $\mu$ for four pinning forces: (a) hard ($c>0$)
cubic pinning force, with $a=c=1/(\pi+\pi^3)$; 
(b)  piecewise linear pinning force, with $a=1/\pi$; 
(c) soft ($c<0$) cubic pinning force, with $a=1/\pi$ and $c=- 1/\pi^3$; 
(d) sine pinning force whose maximum strength is $1/\pi$.
Also shown is the value $\mu_d$ where the coherence jumps from a finite value 
to zero upon decreasing $\mu$.
}
\label{r(mu)}
\end{center}
\end{figure}

The behavior of the coherence as a function of $\mu$ is shown in
Fig.~\ref{r(mu)} for four pinning potentials (for histories where the
effective coupling $u$ is adiabatically increased.) As anticipated on
the basis of the perturbation theory, for a hard pinning force (curve
(a) of Fig.~\ref{r(mu)}) the coherence is a single-valued function of
$\mu$. The system exists in the zero-$r$ IS phase for $\mu<\mu_u$. At
$\mu_u$ there is a continuous transition to the CS phase, with $r$
growing continuously from zero. For soft pinning forces (curves (c),
cubic pinning force, and (d), sine pinning force, of Fig.~\ref{r(mu)})
with $c<0$ the coherence is a multivalued function of $\mu$. In this
case the IS phase is stable up to $\mu_u$ when $\mu$ is ramped up from
below. At $\mu_u$ the coherence jumps discontinuously to the stable
upper branch of the curve corresponding to the CS phase. When $\mu$ is
ramped down from above $\mu_u$, the system remains in the CS phase
down to the lower value $\mu_d$. For this class of pinning forces, the
IS-CS transition is always hysteretic at $F=0$. In the marginal case
of a piecewise linear pinning force (curve (b)), $r$ jumps
discontinuously at the transition, but there is no hysteresis.

\begin{figure}
\begin{center}
\epsfxsize=8cm
\epsfbox{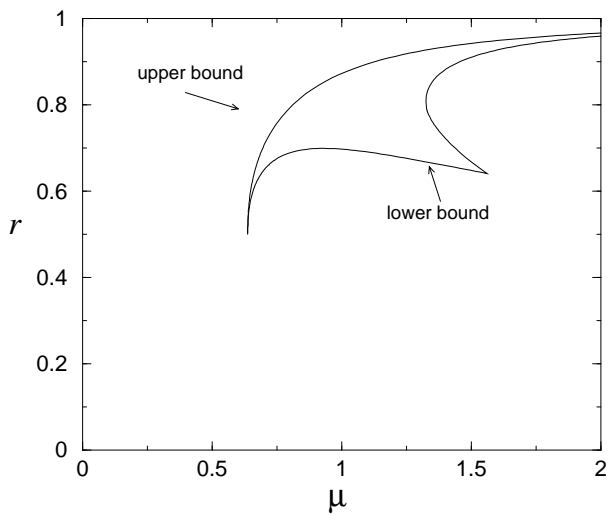}
\caption{Upper and lower bounds for $r(u)$, plotted
as the coherence $r(\mu=u/r)$,
corresponding to the maximal and minimal coherence static metastable
states. The pinning force is taken to be piecewise linear
with $a={1/\pi}$. A
single static coherent solution is obtained only in the limit
$\mu\rightarrow\infty$.}
\label{upperlowerr(mu)}
\end{center}
\end{figure}

The coherence curves shown in Fig.~\ref{r(mu)} correspond to the
metastable states that would result through adiabatically increasing
$u$. As mentioned earlier, for a given $u$, this is the state whose
phases are as close as possible to the global phase $\psi=0$, and
hence is the state with the largest coherence. Thus, the curves shown
in Fig.~\ref{r(mu)} are {\em upper} bounds on the coherence for each
type of pinning force. In order to calculate the lowest possible
coherence at each $u$, one must consider the metastable state whose
phases are as far as possible from the global phase. To obtain this
lower $r(\mu)$ bound analytically is tedious, and we have done so only
for the sawtooth linear case.  This result for the lower bound
is displayed in
Fig.~\ref{upperlowerr(mu)}, along with the upper bound, which, again,
is the relevant state for the histories we consider here.

In addition to determining the transitions between the IS and CS
phases, the non-perturbative treatment at zero drive can also be used
to determine if there is a critical value of $\mu$, $\mu_T$, above
which the depinning threshold vanishes and the system is always
sliding for all $F>0$. We present an outline of the argument here and
relegate the details of the calculation of $\mu_T$ to Appendix A. The
threshold force can be thought of as the largest value of the driving
force at which there still exists a stable static solution to the
equation of motion. All such solutions {\em satisfy the static
self-consistency condition}. For incoherent static solutions, in which
the domains are completely decoupled, this threshold force is simply
the single particle depinning force. For coherent static states the
solution $\delta(\beta)$ is multivalued, but only those metastable
states which satisfy the imaginary part of the self-consistency
condition are acceptable solutions. Consider a system in which there
are multiple metastable static solutions at zero drive. When an
infinitesimal driving force is applied a correspondingly infinitesimal
number of these states becomes unstable as they no longer satisfy the
self-consistency condition. The system remains, however, pinned
provided there still exist other accessible static metastable states.
As the force is further increased, more static states become unstable,
but the system does not depin until the ``last'' of the available
static solutions, that is the one corresponding to the largest value
of $F$ for which a metastable static state exist, becomes
unstable. This value of $F$ defines the depinning threshold.  On the
other hand, if there is a unique metastable static solution at zero
drive, the system will depin immediately upon an infinitesimal
increase of the driving force. Whenever there is a unique solution at
$F=0$, the depinning force is therefore zero.  As shown in Appendix B,
for discontinuous forces there are always a variety of metastable
static states at zero drive for any finite value of $\mu$ (see also
Fig.~\ref{upperlowerr(mu)}), so that $\mu_T=\infty$.  For continuous
pinning forces, there is a finite coupling $\mu_T$ above which there
is a single static state at zero drive and where the threshold force
vanishes.  This is for instance the case for the sinusoidal pinning
force, where the upper and lower bounds of $r(\mu)$ (shown in
Fig.~\ref{r(mu)}) coincide and $\mu_T=\mu_u$.  For a general
continuous pinning force $\mu_T$ is given by
\begin{eqnarray}
\mu_T={\pi|Y^\prime(\pi)|\over\int_0^\pi d \delta\sqrt{1-(Y(\delta)/Y^\prime(\pi))^2}}\;.
\label{mu_T}
\end{eqnarray}

\section{Stability of the Static Incoherent Phase at Nonzero Drive}
\label{IS phase for F nonzero}

We next consider static states in the presence of a finite driving
force, $F\ne 0$, starting with {\em incoherent } static solutions. We
will use a perturbative treatment analogous to that of
Sec.~\ref{Static States} to analyze the limit of stability of the IS
phase against varying $\mu$ and $F$.  For finite $F$, the IS phase can
become unstable to either the coherent static phase {\em or} the
moving phase.  The perturbative analysis described in this section
allows us to establish whether the transition from the IS to CS phase
at finite $F$ is continuous or hysteretic, in much the same way as
done in Sec.~\ref{Perturbation Theory} for $F=0$.  Again we find that
the nature of the transition depends on the type of pinning potential,
but the addition of a driving force changes the shape of the effective
pinning force. This change can, in some cases, change a
continuous IS$\leftrightarrow$CS transition at $F=0$ to a hysteretic
transition at finite $F$.  The value of $F$ above which the CS phase
becomes unstable to a moving state cannot be determined perturbatively
and we defer its calculation to the next section.

The perturbation theory described below is of course only valid for
forces less than the single particle depinning force, $F_{\rm sp}$. This
force is the maximum value of $|Y(x)|$ and is the driving force
required to set in motion a single independent domain. It is hence the
threshold force for an incoherent group of domains.

We will study the stability of the incoherent phase to small changes
in the coherence $r$.  Taking the initial static phase to be
incoherent, the effective coupling $u=\mu r=0$ and the static solution
is obtained by simply balancing the pinning and driving forces. From
Eq.~(\ref{mft_eom1}) (with ${\dot \theta}=0$ and $h=1$) the
noninteracting static solution is $\theta=\beta - Y^{-1}(F)$.  It is
convenient to choose the global phase to be non-zero,
$\psi=-Y^{-1}(F)$, and to work with the deviation ${\tilde
\delta}=\theta-\beta + Y^{-1}(F)$ from the incoherent static solution at a
given $F$.  The static solutions are then given by
\begin{equation} 
0= F - u \sin ({\tilde \delta}+\beta) +  Y({\tilde\delta}+\delta_0)\;,
\label{force_balance_F} 
\end{equation} 
where $\delta_0=\psi=-Y^{-1}(F)$. 
For small $u$ we can expand the pinning force
in powers of ${\tilde\delta}$,
\begin{eqnarray} 
Y_{eff}(\tilde\delta)&\equiv& Y({\tilde\delta}+\delta_0)+ F \\\nonumber
  &=& {\tilde a}(F) {\tilde\delta} + {\tilde b}(F){\tilde\delta}^2 + {\tilde c}(F){\tilde\delta}^3 + \dots \;.
\label{general_pinning_F} 
\end{eqnarray} 
The effective pinning force $Y_{eff}(\tilde\delta)$ has precisely the
same form as $Y(\delta)$ for zero $F$, but the coefficients now depend
on $F$ through $\delta_0=Y^{-1}(F)$. These modified coefficients are
given by
\begin{subequations}
\begin{eqnarray}  
{\tilde a}(F)&=&Y'(\delta_0)\;,\\ 
{\tilde b}(F)&=&Y''(\delta_0)/2\;,\\
{\tilde c}(F)&=&Y'''(\delta_0)/6\;.
\end{eqnarray}
\label{expanded_Y(delta,F)_1}
\end{subequations}
At non-zero drive the coefficient ${\tilde b}(F)$ is always finite,
reflecting the fact that the external drive makes the pinning force
asymmetric about $\delta_0$. The equation for ${\tilde\delta}(\beta)$
is then formally identical to that for $\delta(\beta)$ in the $F=0$
case, with $Y(\delta)\rightarrow Y_{eff}(\tilde\delta)$,
\begin{equation} 
0=- u \sin ({\tilde \delta}+\beta) +  Y_{eff}({\tilde\delta})\;. 
\label{eff_force_balance} 
\end{equation} 
Similarly, the self consistency conditions can be expressed in terms of 
${\tilde \delta}$ as
\begin{equation} 
r ={1\over2\pi}\int_{2\pi}d \beta \cos({\tilde\delta}+\beta)\equiv f(u,F)\;,
\label{self_consistency1aF} 
\end{equation} 
where $r$ is now a function of both $u$ and $F$, and
\begin{equation} 
0 =\int_{2\pi}d \beta \sin({\tilde\delta}+\beta)\;.
\label{self_consistency1bF} 
\end{equation} 
We can now use the results obtained in the zero drive perturbation
theory. The value of $\mu$ at which the IS phase becomes unstable is given by
\begin{equation} 
\mu_u(F)=2 {\tilde a}(F)\;,
\label{mu(F)} 
\end{equation} 
and will
now in general depend on $F$.
Conversely, we can define a critical line $F_u(\mu)$ as the solution of 
$\mu=2 {\tilde a}(F_u)$. 

For drives sufficiently small that the system remains pinned at the
instability line, the form of the onset of coherence near
$\mu_u(F)$ can be determined by looking for a solution to
Eq.~(\ref{self_consistency1aF}) in the form of a power series,
\begin{equation} 
f(u,F)=r_1(F)u+r_2(F)u^2+r_3(F)u^3+\dots\;.
\label{f(u,F)} 
\end{equation} 
As usual in such calculations, we expect the nature of the instability
to depend on the signs of the coefficients.
The coefficients $r_1(F)$, $r_2(F)$ and $r_3(F)$ are given by 
Eq.~(\ref{expanded_f(u)_2}) with $a$, $b$, $c$ replaced by 
${\tilde a}(F)$, ${\tilde b}(F)$, ${\tilde c}(F)$, giving
\begin{subequations}
\begin{eqnarray}  
r_1(F)&=&{1\over2 {\tilde a}(F)}\;,\\ 
r_2(F)&=&0\\
r_3(F)&=&{3\over8}\left({{\tilde a}(F){\tilde c}(F)-
2{\tilde b}(F)^2\over {\tilde a}(F)^5}\right)\;.\label{f(u)_2cF}
\label{expanded_f(u)_F}
\end{eqnarray}
\end{subequations}
Thus, the form of $r(\mu,F)$ near $\mu_u(F)$ is:
\begin{equation} 
r(\mu,F)={1\over\mu^3}\left({\mu-\mu_u(F)\over r_3(F)\mu_u(F)}\right)^{1/2}
\label{pert_r(mu,F)1}\;. 
\end{equation} 
As for the case of $F=0$, the behavior is controlled by the sign of
the coefficient $r_3(F)$ of the cubic term in Eq.~(\ref{f(u,F)}). If
$r_3(F)>0$ the coherence grows as $\sim(\mu-\mu_u(F))^{1/2}$ with
increasing $\mu$, indicating that the $r$ versus $\mu$ curve is
continuous. Conversely, if $r_3(F)<0$ the coherence grows with {\em
decreasing} $\mu$ as $\sim(\mu_u(F)-\mu)^{1/2}$, and the $r$ versus
$\mu$ curve is hysteretic. One important complication is that for
finite $F$ the coefficient $r_3$ can change sign as a function of $F$
for a given pinning force. As a result the
transition between coherent and incoherent static states can change
from continuous to hysteretic above a characteristic force $F_{\rm h}$
defined by the solution of $r_3(F_{\rm h})=0$.

We now specifically apply these general results to the three classes
of pinning forces (linear, hard, and soft.) Again, these are of the
general form
\begin{equation}
Y(x) = -a x -cx^3\;, -\pi \leq x  \leq \pi\;,
\label{Y(x)}
\end{equation}
with $a>0$. The three classes have $c$ zero, positive and negative,
respectively.

\subsection{Piecewise Linear Pinning Force ($c=0$)}
\label{Sawtooth Pinning Force}

\begin{figure} 
\begin{center}
\epsfxsize=5cm
\epsfbox{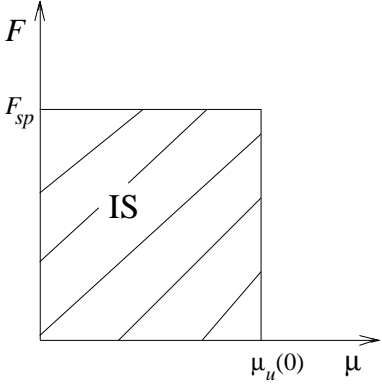}
\caption{
The region of stability of the IS phase for a piecewise linear pinning
force. The single particle depinning force $F_{\rm sp}$ and the
coupling strength $\mu_u$ for instability to the coherent or moving
states are also indicated. }
\label{Sawtooth_IS_stability}  
\end{center}
\end{figure}

For the piecewise linear pinning force of Fig.~\ref{potentials}(e),
where $Y(\delta)$ is given by Eq.~(\ref{Y(x)}) with $c=0$, we simply
have ${\tilde a}(F)=a$ and ${\tilde b}(F)={\tilde c}(F)=0$. In this
case $\mu_u(F)=\mu_u(0)$, independent of $F$. In fact we will show in
Section~\ref{Piecewise Linear Pinning Force Phase Diagram} that the
coherence $r(\mu)$ of the entire static state is independent of $F$
for all values of $\mu$, whenever the system is pinned. The IS
phase is stable for $\mu<\mu_u=2a$ and $F<F_{\rm sp}=a\pi$. This region is
shown in Fig.~\ref{Sawtooth_IS_stability}.

\subsection{Hard Cubic Pinning Force ($c>0$)}
\label{Hard Cubic Pinning Force}

In Fig.~{\ref{mu_c(F)} we show the region of stability of the IS phase
for the hard pinning force of Fig.~\ref{potentials}(d).  In this
example, the maximum pinning force from Eq.~(\ref{Y(x)}) gives the
single particle depinning threshold as $F_{\rm sp}= a \pi +
c\pi^3$. When the coupling $\mu$ is ramped up adiabatically with
constant $F<F_{\rm sp}$, the IS state becomes unstable at a value
$\mu_u$ given by (see Eq.~(\ref{mu(F)})),
\begin{equation}
\mu_u(F)=2\big[a-3c\delta_0^2(F)\big]\;.
\end{equation}
For the hard cubic potential this result can be inverted analytically
to obtain the boundary $F_u(\mu)$ of the IS state shown in Fig.~\ref{mu_c(F)},
with the result
\begin{equation} 
F_u(\mu)=\frac{2\mu_u+\mu}{6}\sqrt{\frac{\mu-\mu_u}{6|c|}}\;.
\label{F_u^h_cubic}
\end{equation} 
The maximum value of $\mu$ for which the IS state is stable
is $\mu^*$, where $\mu^*$ is found
by the intersection of the IS depinning curve and the $F_u(\mu^*)$ curve.
Its value is
\begin{equation}
\mu^*=\mu_u+6c\pi^2\;.
\label{mustar}
\end{equation}
Note that if the system is prepared in the IS state at $\mu>\mu_u$,
then a transition to a coherent state can be achieved by {\em
decreasing} $F$. This is because decreasing $F$ allows the domains to
relax back toward the minima of their pinning potentials, where the
pinning force (determined by the curvature of the potential) is
smaller and hence the coherence can increase.

\begin{figure} 
\begin{center}
\epsfxsize=5cm
\epsfbox{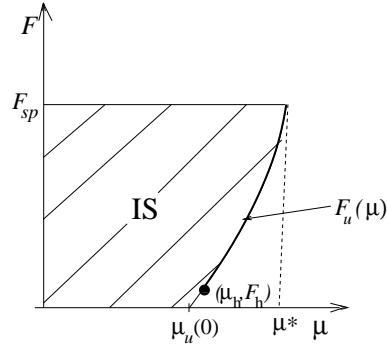}
\caption{A plot of the region of stability of the incoherent 
static phase for a hard cubic pinning force. The nature of the
instability along the $F_u(\mu)$ curve is indicated by the thickness
of the bounding curve on the right. For $\mu>\mu_{\rm h}$ ($F>F_{\rm h}$), the
transition is hysteretic, while for smaller couplings (or small, fixed
 driving force for varying couplings) the transition is continuous. }
\label{mu_c(F)}  
\end{center}
\end{figure} 

In the case of the hard cubic pinning force the coefficient $r_3(F)$
can change sign as a function of $F$. For small $F$, $r_3(F)>0$ and
the transition from the IS to a coherent static phase is continuous. Above a
critical value $F_{\rm h}$ defined by $r_3(F_h)=0$ the transition becomes
hysteretic.  The force $F_{\rm h}$ is given by
\begin{equation} 
F_{\rm h}={16\over15^{3/2}}\sqrt{a^3\over c} \;,
\label{F_h_cubic}
\end{equation} 
and is small compared with $F_{\rm sp}$ for the potential shapes and parameters
we have considered. For $a=c=1/(\pi+\pi^3)$, 
we find $F_{\rm sp}=1$, $F_{\rm h}\approx 0.008$ and $\mu_{\rm h}\approx 1.2\mu_u$. 

\subsection{Soft Cubic Pinning Force ($c<0$)}
\label{Soft Cubic Pinning Force}

Soft cubic pinning forces given by Eq.~(\ref{Y(x)}) with $c$ negative,
can be divided into two classes: (i) forces
that are monotonic
functions of the phase within each period, as plotted in
Fig.~\ref{potentials}(a),
and (ii) those
that reach their maximum (minimum) within a given period and turn over,
as plotted in Figs.~\ref{potentials}(b) and
\ref{potentials}(c).
Holding $\mu$ constant, the incoherent static state 
becomes unstable upon increasing $F$ to $F_u(\mu)$, with
\begin{equation} 
F_u(\mu)\equiv \frac{2\mu_u+\mu}{6}\sqrt{\frac{\mu_u-\mu}{6|c|}}\;,
\label{F_u_soft}
\end{equation}
unless the single particle depinning force is first reached.  For
pinning forces in class (i) the value $\mu^*$ where
$F_u(\mu^*)=F_{\rm sp}$ is positive and the region of stability of the
incoherent static state is of the type shown schematically in
Fig.~\ref{soft_mu_c(F)}(a). For pinning forces in class (ii) (for
pinning forces with only cubic terms, this class is given by
$|c|\geq1/(3\pi^2)$), it can be shown that $\mu^*= 0$. The single
particle depinning transition is always preempted. 
Here, the region of stability of the incoherent state is determined 
by $F_u(\mu)$ for all values of $\mu$, as shown in
Fig.~\ref{soft_mu_c(F)}(b). 

\begin{figure}
\begin{center}
\epsfxsize=9cm
\epsfbox{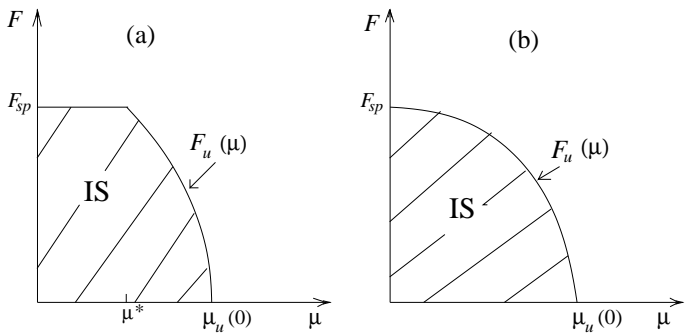}
\caption{Sketches of the region of stability of the incoherent static 
phase for a soft cubic pinning force. Figure (a) corresponds to a
pinning force of type (a) that does not turn over (is monotonic) in
each repeated interval. Figure (b) corresponds to a pinning force of
type (b) that are non-monotonic in each period.}
\label{soft_mu_c(F)}  
\end{center}
\end{figure} 

For a soft cubic pinning force $r_3(F)$ is negative for all $F$
and the transition from the incoherent to a coherent state is always
hysteretic.

\section{Nonequilibrium Phase Diagrams in the $\mu$-$F$ plane.}
\label{nonzero F COV}

In this section we present the nonequilibrium phase diagrams in the
$\mu$-$F$ plane for the various pinning forces introduced in
Fig.~\ref{potentials}.  The phase diagrams are based upon both
analytical results and numerical computations. The analytical bounds on the
stability of the static phases are based on the previous sections'
results for the incoherent static phase and calculations for the
coherent static phase whose details are presented in the appendices.
Numerical integration of the equations of motion is used to determine
the boundaries of the moving phases: by starting from the moving phase
and decreasing $F$ or $\mu$, the repinning curves can be found. 
Of special interest is the nature of the depinning
transition obtained when the applied force $F$ is varied at constant
$\mu$.  The curves of mean velocity as a function of driving force
correspond to the IV characteristics of physical systems, such as CDWs
and vortex lattices. Our focus is on classifying models or parameter
ranges for which the depinning transition is continuous or hysteretic.
In general, for each of the pinning forces we consider, the depinning
transition appears to be continuous with a unique depinning threshold
at large $\mu$, where the system is more rigid. In contrast, the
velocity-force curves generally exhibit macroscopic hysteresis at
small values of $\mu$, where the system is more likely to display
plastic effects.

\subsection{Piecewise Linear Pinning Force}
\label{Piecewise Linear Pinning Force Phase Diagram}

In Sec.~\ref{Sawtooth Pinning Force}, perturbation theory was employed
to study the transition between incoherent and coherent static phases
for the piecewise linear pinning force.
It was found that when the coupling strength $\mu$ is changed at fixed
$F$ within the pinned region of the phase diagram this transition is always
discontinuous, although not hysteretic. Furthermore, the critical
value of $\mu$ where the transition takes places appears to be
independent of the driving force.  Here we show that this remains true
in a complete calculation. We also calculate the depinning threshold
exactly by determining the limit of stability of the static phases.
For the 
piecewise linear pinning force (i.e., Eq.~(\ref{Y(x)}) with $c=0$),
the force balance equation in the static state is
\begin{equation} 
0= F - u \sin (\delta+\beta) -  a\delta\;,
\label{sawtooth_force_balance_F1} 
\end{equation} 
where $-\pi\leq\delta\leq\pi$ and we have chosen $\psi=0$.
Letting $\delta={\tilde \delta}+F/a$ and $\beta=\alpha-F/a$,
Eq. (\ref{sawtooth_force_balance_F1}) can be written as 
\begin{equation} 
0= - u \sin ({\tilde \delta}+\alpha) -  a {\tilde \delta}\;,
\label{sawtooth_force_balance_F2} 
\end{equation} 
with $-\pi-F/a \leq {\tilde \delta} \leq \pi-F/a$. 
It is apparent from Eq.~(\ref{sawtooth_force_balance_F2}) 
that ${\tilde \delta}$ is a function only of $\alpha$ and $u$ and
does not depend on $F$ explicitly. 
The real part of the self-consistency condition that determines
the coherence $r$ becomes
\begin{equation} 
r ={1\over2\pi}\int_{-\pi}^{\pi}d \alpha \cos({\tilde \delta}+\alpha)\;,
\label{self_consistency_x_1} 
\end{equation} 
and clearly $r(u,F)=r(u,F=0)$. Thus, the coherence of the static state
is independent of $F$. The line separating the incoherent and coherent
static phases is a vertical line at $\mu=\mu_u=2a$ in the $\mu-F$ plane, as
shown in Fig.~\ref{linear}. The IS-CS transition is discontinuous and
non-hysteretic at all values $F$ where the static phases are stable.   
When the force is
ramped up adiabatically at fixed $\mu<\mu_u$ from the IS phase where $r=0$, 
the
system depins at the single particle depinning force
$F_{\rm sp}=a\pi$. For $\mu > \mu_u$ the system is in the CS phase, where
the coherence is non-zero and ${\tilde \delta}$
is a multivalued function of $\alpha$. As discussed in Sec.~\ref{COV},
there are many static metastable states available to the system for a 
fixed value of $u$.
We relabel the metastable states and denote each state by 
a ${\hat \delta}_i(\alpha,u)$
which is a single valued, but generally discontinuous, function of $\alpha$. 
Each ${\hat \delta}_i$ must satisfy the imaginary part of the 
self-consistency condition which using 
Eq.~(\ref{sawtooth_force_balance_F2}) can be rewritten as
\begin{equation} 
0 =\int_{-\pi}^{\pi}d \alpha {\hat \delta}_i(u,\alpha)\;.
\label{self_consistency_x_2} 
\end{equation} 
This implies that the acceptable ${\hat \delta}_i$'s are odd functions
of $\alpha$. In addition, each static metastable solution must lie
within the upper and lower bounds, ${\tilde \delta}_u(F)\equiv
\pi-F/a$ and ${\tilde \delta}_l(F)\equiv -\pi-F/a$. As $F$ is
increased, the value of the upper bound decreases, reducing the number
of allowed ${\hat \delta}_i$'s, until at $F=F_{\uparrow}^c(\mu)$ only
one solution remains. This special state, ${\hat \delta}(\alpha,u)$,
is equivalent to the one that would be obtained through adiabatically
increasing $u$. The associated $r(\mu)$ curve is shown in
Fig.~\ref{r(mu)} for $a=1/\pi$. The value of $F_{\uparrow}^c(u)$
\cite{F_up_footnote} is given by ${\hat
\delta}_{max}(u)=\pi-F_{\uparrow}^c/a$.  For $u\leq a\pi/2$ we find
from Eq.~(\ref{sawtooth_force_balance_F2}) ${\tilde
\delta}_{max}(u)=u/a$ which gives
\begin{equation} 
F_{\uparrow}^c(u)=\pi-u/a\;, \hspace{0.5in} u\leq a\pi/2\;.
\label{F_T(u)1} 
\end{equation} 
For $u\geq a\pi/2$ ${{\hat \delta}_{max}}$ is defined implicitly 
by $a{\hat \delta}_{max}=u \sin({\hat \delta}_{max})$ and $F_\uparrow^c$ is given by
\begin{equation} 
a\pi-F_{\uparrow}^c(\mu)=u \sin\big[\pi F_{\uparrow}^c(\mu)-u/a\big]\;,  \hspace{0.1in} u\geq a\pi/2\;.
\label{F_T(u)2} 
\end{equation} 
It is then possible to calculate $F_{\uparrow}^c(\mu)$ using the
expression for $r(u)$ given in Eq.~(\ref{sawtooth_Y_r(u)}).  
The resulting phase diagram is shown in Fig.~\ref{linear}.

\begin{figure}
\begin{center}
\epsfxsize=8cm
\epsfbox{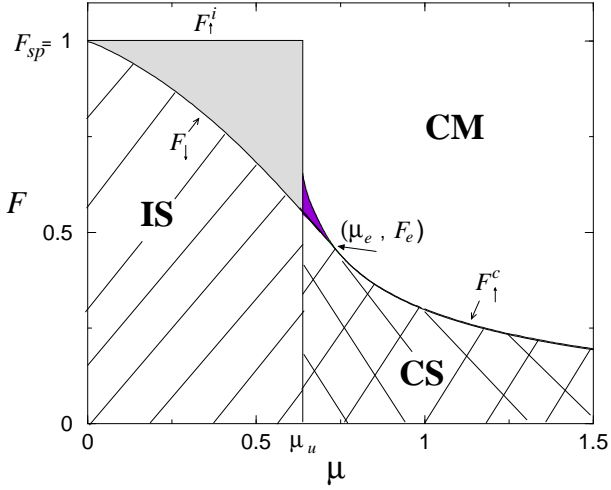}
\end{center}
\caption{ \label{linear} Phase diagram for the piecewise linear pinning force,
$Y(x)=-x/\pi$ (see Fig. (\ref{potentials}b)).  The lightly shaded
portion is the coexistence region of the IS and CM phase ($\mu<\mu_u$)
and the smaller, darkly shaded region, is where the CS and CM phases
coexist ($\mu_u<\mu<\mu_e$).  The depinning lines
$F_{\uparrow}^i=F_{\rm sp}$ and $F_\uparrow^c$ have been obtained
analytically and confirmed by numerics.  The boundary $F_\downarrow$
where the system repins was obtained numerically. The point
$(\mu_e,F_e)$ marks where the static-moving transition changes from
hysteretic to continuous.  The boundary between the IS and CS phases
is $F$ independent and lies at $\mu=\mu_u$.}
\end{figure}

For $\mu<\mu_u$ the static phase is incoherent and the depinning
transition is hysteretic in both $v$ and $r$, as shown in the top two
frames of Fig.~\ref{vFlinear}.  The system depins at $F_{\rm sp}$ when the
drive is ramped up adiabatically from the static phase, but repins at
the lower force $F_\downarrow$ when the force is ramped back down from
the sliding state.  The line $F_\downarrow$ has been obtained by
numerical simulation of the mean field model. The numerics have also
revealed that a small region of hysteresis persists for $\mu>\mu_u$,
although the static phase is coherent here.  The behavior of $v$ and
$r$ in this region is shown in the two mid frames of
Fig.~\ref{vFlinear}.  Finally, for $\mu>\mu_e$ (where $\mu_e$ is the
value of the coupling above which the static-moving transition is
elastic in nature) the depinning is continuous, as shown in the bottom
frames of Fig.~\ref{vFlinear}. The values of $\mu_e$ and $F_e$ are defined via
$F_\uparrow^c(\mu_e)=F_\downarrow(\mu_e)=F_e$. Finally, the depinning
threshold $F_{\uparrow}^{c}$ is nonzero and finite for all $\mu$,
i.e., $\mu_T=\infty$. This is a general property of discontinuous
pinning forces, to be contrasted with the behavior observed for
continuous pinning forces, such as the sinusoidal one studied by
Strogatz and collaborators. \cite{Strogatz}

\begin{figure}
\begin{center}
\epsfxsize=8cm
\epsfbox{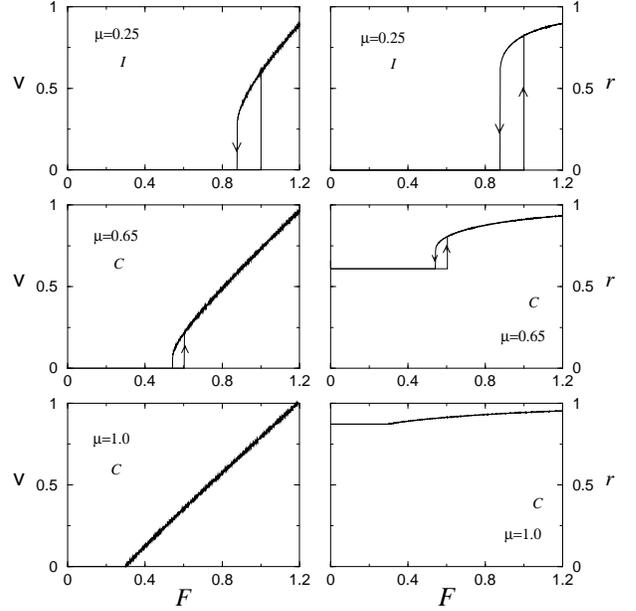}
\end{center}
\caption{Mean velocity $v$ and coherence $r$ as functions of the driving force 
$F$ for the piecewise linear pinning force.  The curves are obtained
numerically by first ramping $F$ from zero to a value well within the
sliding state ($F=1.2$), and then decreasing $F$ back down to zero, while
holding $\mu$ constant.  The top frames show the behavior for $\mu=0.25$,
where the initial static state is incoherent: this state starts sliding
at the single-particle depinning force $F_{\rm sp}=1$ and repins at a lower
force $F\approx 0.88$.
The mid frames display the results for an initially coherent static state
($\mu=0.64>\mu_u$), which still displays hysteresis, both in $v$ and
$r$.  The bottom
frames are for $\mu=1.0$, which has an initial coherent state
and undergoes continuous depinning.  }
\label{vFlinear}
\end{figure}

Before closing this subsection, we must address the possibility an of 
incoherent moving (IM) phase.  Strogatz and collaborators \cite{Strogatz} found that an IM 
phase is always unstable for a sinusoidal pinning potential.
It can be shown that this remains true for other continuous pinning forces.  
The situation is less clear for discontinuous 
pinning forces.  In Appendix D we present the details of a short time ($t=0$) 
stability analysis for the IM phase for any $Y(x)$.  This analysis will
tell us something about the 
long time, steady state limit, provided 
$r(t)$ is a monotonic function of time. This analysis predicts
a range of stability for the IM phase for discontinuous pinning forces,
provided the jump discontinuity at $x=\pi$ is taken into account when 
preparing the system.  However, simulations show that $r(t)$ is in general
not monotonic and that the strength of the perturbation needs to be decreased 
with system size in order to observe the IM phase, suggesting that the 
perturbative short-time analysis is simply
not valid in this case.  Finally, 
if a narrow distribution of pinning strengths $h$ is introduced, we find numerically 
the IM phase 
to be unstable.  Given these numerical findings, we believe that the IM 
phase is generally unstable in mean field theory.  

\subsection{Hard Cubic Pinning Force}
\label{Hard Cubic Static Phase Diagram}

The phase diagram for a hard cubic
pinning force, given by Eq.~(\ref{Y(x)}) with $c>0$ 
(see Fig.~\ref{potentials}(d))
is shown in
Fig.~\ref{hard} for $a=c=1/(\pi+\pi^3)$.

\begin{figure}
\begin{center}
\epsfxsize=8cm
\epsfbox{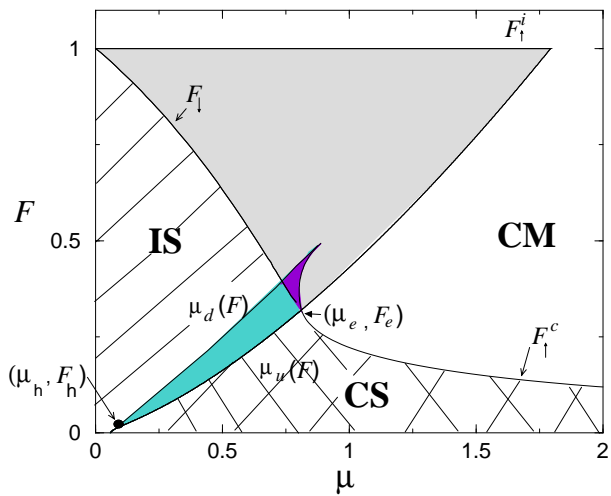}
\end{center}
\caption{ \label{hard} Phase diagram in the coupling-drive ($\mu$-$F$) 
plane for a hard cubic pinning force 
 of the type shown in Fig. (\ref{potentials}a). The form of the pinning force
$Y(x)$ is given by Eq. (\ref{Y(x)}), with  $a=c=1/(\pi+\pi^3)$. 
The regions of IS-CM, CS-CM and IS-CS-CM coexistence are shown in light, 
medium and dark gray, respectively. The incoherent and coherent depinning 
lines are denoted by $F_{\uparrow}^i$ and $F_{\uparrow}^c$ respectively. 
The repinning line is denoted by $F_{\downarrow}$. The coherent depinning line and the repinning line join at $(\mu_e, F_e)$. Beyond this point the 
static-moving transition is continuous. The curves $\mu_u(F)$ and 
$\mu_d(F)$ are the values of the coupling at which the static system 
makes the transition to and from finite coherence states,
respectively. There curves join at $(\mu_{\rm h}, F_{\rm h})$ where the IS-CS 
transition becomes continuous.}
\end{figure}

Though the general topology is similar to that of the phase diagram
for the piecewise linear force, the history dependence is
significantly more complicated.  A first difference is that the
transition between the IS and CS phases is now continuous for $F<F_h$,
with $F_{\rm h}$ given by Eq.~(\ref{F_h_cubic}), and hysteretic for
$F>F_{\rm h}$. For the parameter values displayed in Fig.~\ref{hard} the
value of $F_{\rm h}$ is very small, but still finite.  A second new feature
of the phase diagram is the presence of a small region (darkest gray
in Fig.~\ref{hard}) where all three phases coexist.

The strong history dependence is manifested in the macroscopic
response and includes reentrant behavior for fixed $\mu$ or $F$
histories.  The mean velocity and coherence are plotted as a function
of (increasing, then decreasing) driving force for a few typical
values of $\mu$ in Fig.~\ref{vfhard1}. The pinning force is given by
$Y(x)=-(x+x^3)/(\pi+\pi^3)$. The top frames show a simple hysteretic
depinning transition for a system prepared in the incoherent static
state at $F=0$, similar to that seen for a linear pinning force.  The
middle row of frames display the more complicated history that results
when the system is prepared in a coherent static state at $F=0$, with
$\mu=0.5$.  The velocity shows a single hysteresis loop, but the plot
of coherence $r$ shows first a decrease and then a jump to the
incoherent state as the force is increased, followed by a jump back to
a finite value when bulk depinning takes place.  In this case both the
regions of IS-CS and IS-CM coexistence are crossed when $F$ is ramped
up.  The IS-CS transition occurs as the phases are pushed away from
their zero-force minima to regions of the pinning potential with
higher curvature, which makes the coherent state unstable.  Upon
decreasing the force, both the coherence and the velocity jump back to
zero, then the coherence increases again as the force is
decreased. The jumps in coherence when $F$ is ramped down occur at
values of $F$ different from those where the coherence jumps during
the ramp up.  For rather specific values of $\mu$, even more baroque
histories can be found by crossing the three-phase coexistence
regions. An example is shown in the last row of frames in
Fig.~\ref{vfhard1}, where $\mu=0.76$.  Here, the sequence is
CS$\rightarrow$IS$\rightarrow$CM$\rightarrow$CS, which skips the IS
phase on decreasing $F$. Note that the velocity vs.\ drive force curve
is relatively unremarkable, showing simple hysteresis in this
case. The coherence history is more complicated.

\begin{figure}
\begin{center}
\epsfxsize=8cm
\epsfbox{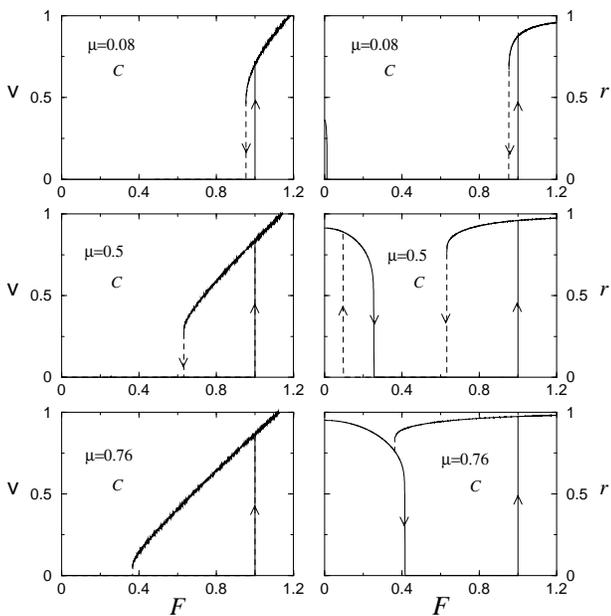}
\end{center}
\caption{\label{vfhard1}
Mean velocity and coherence versus force for the hard potential and
various values of $\mu$. Solid lines are used to display the response
obtained when $F$ is ramped up from zero, while dashed lines show the
jumps in $v$ and $r$ when ramping $F$ back down. The top frames show
the hysteretic depinning of a system prepared in the IS phase.  For
$\mu=0.5$ (mid frames) the system is initially in a coherent ($r\not=0$)
static state at $F=0$. As the force is ramped up, the system first
crosses the boundary from the CS to the IS phase, where $r$ jumps
discontinuously from its initial finite value to zero, while the
system remains pinned ($v=0$).  At a higher force the system depins by
crossing the boundary from the IS to the CM phase and $r$ jumps from
zero to a large finite value. The subsequent ramping down of the field
goes through this sequence of phases in reverse order, but the jumps
occur at distinct values of $F$. The bottom frames describes the
complex response that takes place along a path that crosses the dark
region of three-phase coexistence. See the text for further description.}
\end{figure}

Another interesting feature of the phase diagram for the hard cubic
pinning force is that at constant $\mu$, a portion of the moving phase
lies {\em between} the incoherent and coherent static phases. This
suggests the possibility of re-entrance in the depinning transition
for $\mu>\mu_e$. It is not, however, straightforward to prepare the system 
in the lightly shaded portion of the phase diagram
where IS and CM phases coexist and $\mu > \mu_e$
The static solution must either be
created ``by hand'' at that location $(\mu,F)$ in phase space or the
system can be prepared in the IS phase {\em at a lower value of $\mu$}
and the coupling can then be ramped up to the relevant value
$\mu>\mu_e$.  Both the difficulty of preparing the system in the
re-entrant state and the re-entrance for a
specially prepared state are displayed in
Fig.~\ref{vFhard_mu=1.25}. Here both sets of curves correspond to the
same value of the coupling strength, $\mu=1.25$. In the top pair of
curves the system is prepared in the coherent state at $F=0$. As the
force is ramped up adiabatically, the system depins continuously at
$F^c_\uparrow$, where both velocity and coherence change smoothly,
with $r$ rapidly approaching its limiting value, $r=1$.  The
coexistence region is never accessed in this case.  In the bottom set
of figures, the system is prepared in an incoherent static state at
finite $F$, deep inside the coexistence region.  The system is then
observed to {\em depin} as the force is {\em ramped down} at constant
$\mu$ across the boundary between the coexistence region and the CM
phase. Simultaneously, the coherence jumps from zero to a large finite
value.  Upon further ramping down $F$, the system repins again
continuously at $F^c_\uparrow$.

\begin{figure}
\begin{center}
\epsfxsize=8cm
\epsfbox{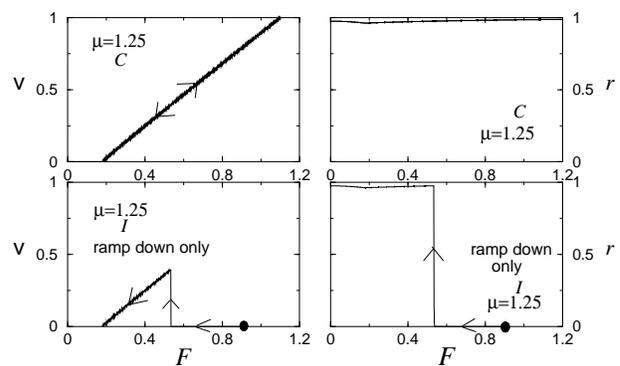}
\end{center}
\caption{Both sets of figures show the behavior of velocity and coherence  for
$\mu=1.25$, but for different initial states.  The top frames are obtained by
preparing the system in a coherent state at $\mu=1.25$ and $F=0$,
and ramping $F$ up to a value above $F^i_\uparrow$, and then back down to zero.
In this case the depinning is continuous. The bottom frames
are obtained by preparing the system 
in an incoherent state at $\mu=1.25$ and $F=0.9$,
inside the lightly shaded area of coexistence of CS and CM phases,
and then ramping the force down to zero. Note the {\em depinning upon
decreasing force} in this case and the subsequent repinning.}
\label{vFhard_mu=1.25}
\end{figure}

\subsection{Soft Cubic Pinning Force}
\label{Soft Cubic Static Phase Diagram}

We distinguish three types of soft cubic pinning forces given by
Eq.~(\ref{Y(x)}) with  $c<0$.
These pinning forces and
corresponding potentials are shown in Fig.~{\ref{potentials}: (a)
forces that are monotonic over the entire period and do not turn over
in the interval $[-\pi,\pi]$; (b) forces that are nonmonotonic over
the period and do turn over in the interval $[-\pi,\pi]$, but are
discontinuous; and (c) continuous forces, which are obviously
nonmonotonic.  
The phase diagrams for these potentials exhibit qualitative differences
as compared to those discussed so far.
Specifically, the CS region at non-zero $F$ may or may not extend to 
$\mu=\infty$ and may not even exist. For most potentials, however, we
do find a non-trivial coherent static phase. The only exception is the 
case of a sinusoidal pinning force studied previously by Strogatz and collaborators,
\cite{Strogatz} where the CS state is unstable. 

For monotonic pinning forces, (a), the boundaries $F^i_\uparrow$ and
$F_{\rm sp}$ intersect at a finite positive value $\mu_*$ of $\mu$, given
by Eq.~(\ref{mustar}) (see Sec.~\ref{Soft Cubic Pinning Force} for a
full discussion).  This results in a portion of the
depinning boundary being horizontal on the $\mu$-$F$ plane,
as $F^i_\uparrow=F_{\rm sp}$ for $\mu\leq\mu_*$, as
shown in Fig.~\ref{soft1}.  In contrast, if the pinning force is
nonmonotonic, (b) or (c), and reaches its maximum within the period,
then $\mu*=0$ and the phase boundary has no horizontal portion.
This behavior is shown in Fig.~\ref{soft3} for a
nonmonotonic, but discontinuous pinning force.

\begin{figure}
\begin{center}
\epsfxsize=8cm
\epsfbox{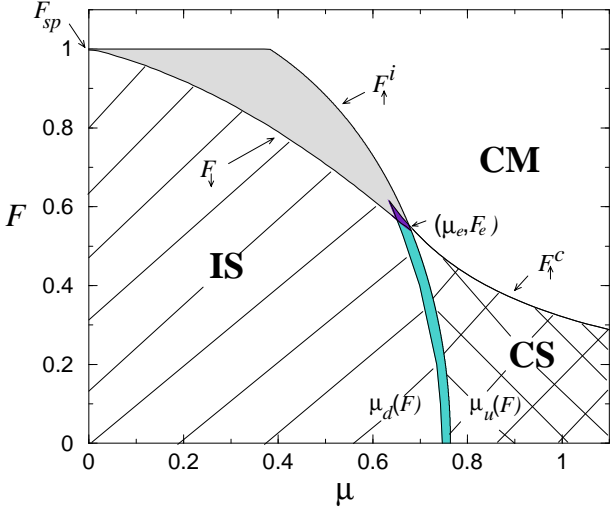}
\end{center}
\caption{ \label{soft1}
Phase diagram in the coupling-drive ($\mu$-$F$) plane for a soft
monotonic cubic pinning force of the type shown in
Fig.~\ref{potentials}(a). The pinning force $Y(x)$ is given by
Eq. (\ref{Y(x)}) with $a=6/(5\pi)$ and $c=-1/5\pi^3$. The regions of
IS-CM, CS-CM and IS-CS-CM phase coexistence are shown in light, medium
and dark gray respectively. The lines $F_{\uparrow}^i$ and
$F_{\uparrow}^c$ are the forces at which the system depins upon
increasing the drive from the incoherent and coherent static states
respectively. The line $F_{\downarrow}$ is the force at which a moving
system stops upon lowering the drive. The coherent depinning line and
the repinning line join up at $(\mu_e,F_e)$ and the static-moving
transition becomes continuous. The curves $\mu_u(F)$ and $\mu_d(F)$
are the values of the coupling at which the static system makes the
transition to and from finite coherence states respectively.}
\end{figure}

The results for pinning forces of type (c), that are continuous (and
therefore must be non-monotonic) have two important features: $\mu*=0$
and $\mu_T$ is finite. These features imply, respectively, that there
is no horizontal portion to the CS depinning curve and that the system
slides at arbitrarily small force whenever the coupling is large,
i.e., when $\mu\geq\mu_T$. The typical phase diagram for a pinning
force of this type is shown in Fig.~\ref{soft4}.
Figure~\ref{soft4_vfrf} shows sample $v(F)$ and $r(F)$ plots for this case.  

\begin{figure}
\begin{center}
\epsfxsize=8cm
\epsfbox{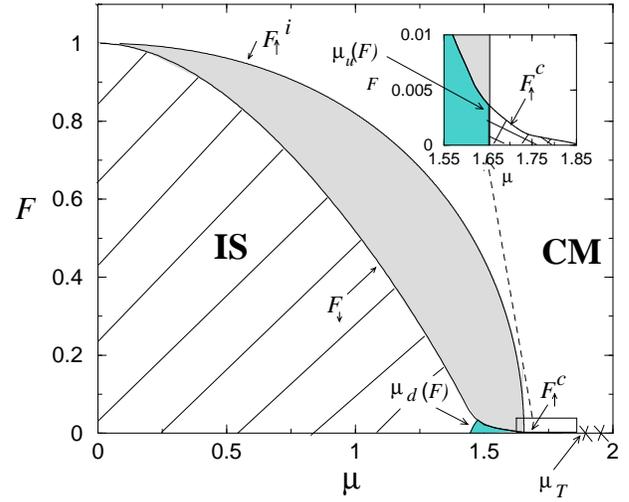}
\end{center}
\caption{\label{soft4}
Phase diagram in the coupling-drive ($\mu$-$F$) 
plane for a soft cubic pinning force of the type shown in
Fig.~\ref{potentials}(c). The pinning force $Y(x)$ is given by
Eq.~(\ref{Y(x)}) with $a=3\sqrt{3}/(2\pi)$ and
$c=-3\sqrt{3}/(2\pi^3)$. This choice of parameters gives a non-monotonic
and {\em continuous} pinning force: the results are to be compared
with $Y(x)=-\sin(x)$, another non-monotonic and continuous force. 
The region of IS-CM coexistence is shown in
light gray, while the IS-CS coexistence is shown in medium gray. 
The lines $F_{\uparrow}^i$ and 
$F_{\uparrow}^c$ are the forces at which the system depins upon 
increasing the drive from the incoherent and coherent static states 
respectively. The line $F_{\downarrow}$ is the force at which a 
moving system stops upon lowering the drive. The CS region is very small 
for these values of parameters, corresponding to 
$\mu_T\approx 1.85$ and $\mu_d(0)\approx 1.44$, 
and it has been magnified in the inset. The CS phase does not 
exist at finite $F$ for coupling larger than $\mu_T$. 
Shown within this inset is the point $(\mu_e,F_e)$ where the 
$F_{\uparrow}^c$ and $F_{\downarrow}$ lines join and the static-moving 
transition ceases to be hysteretic. The curves $\mu_u(F)$ 
(only visible within the inset) and 
$\mu_d(F)$ are the values of the coupling at which the static system 
makes the transition to and from finite coherence states
respectively.}
\end{figure}

\begin{figure}
\begin{center}
\epsfxsize=8cm
\epsfbox{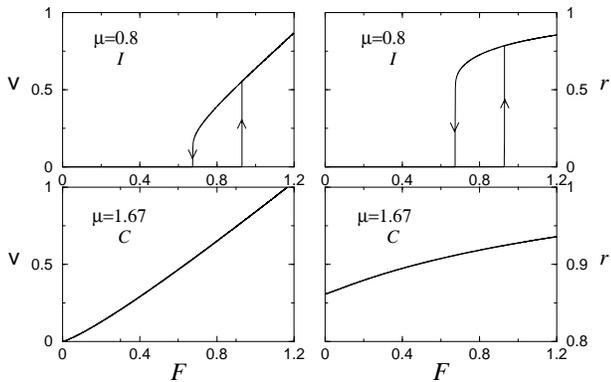}
\end{center}
\caption{\label{soft4_vfrf} 
Mean velocity and coherence, obtained from numerical calculations,
for a continuous cubic pinning potential and parameter values given
in Fig.~\ref{soft4}.  The top frames record
the hysteretic response of a system prepared in the incoherent state
at $\mu=0.8$ and $F=0$, while the bottom frames show the continuous
$F=0$ depinning of system prepared in the coherent state at $\mu=1.67$.}
\end{figure}

At finite drive the CS phase does not extend beyond $\mu=1.84$. For 
values of the coupling between $\mu_d$ and $\mu_T$ the CS phase exists 
at finite drive, albeit only for very small values of 
$F<F_\uparrow^c(\mu)\ll 1$. This small region of the phase diagram in 
Fig.~\ref{soft4} is magnified and shown in the inset. It is interesting 
to compare the these 
results with those obtained by Strogatz and collaborators \cite{Strogatz} for another 
continuous pinning force, namely $Y(x)=-\sin(x)$. The corresponding 
phase diagram is shown in Fig.~\ref{sine}. In this case $\mu_u=\mu_T=2$ 
and, more significantly, $F_\uparrow^c(\mu)=0$. This means that the CS 
phase {\em never} exists at finite $F$. Thus, it seems 
that sinusoidal pinning forces are a special class of more general 
continuous pinning forces in that they never allow the possibility 
of a CS phase at finite drive. This difference, while important 
qualitatively, may not be quantitatively significant given that 
$F_\uparrow^c(\mu)$ is always very small.

\begin{figure}
\begin{center}
\epsfxsize=8cm
\epsfbox{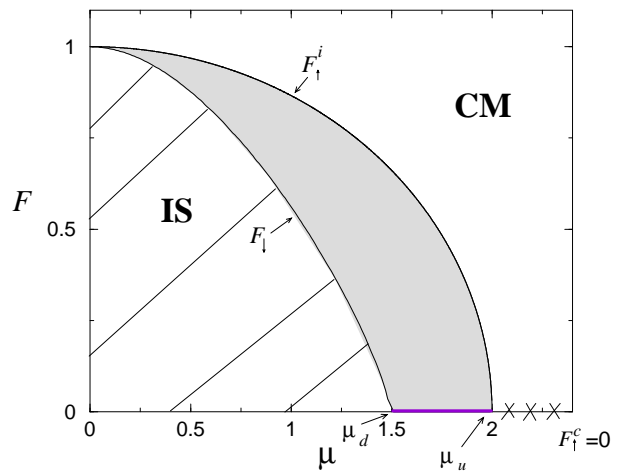}
\end{center}
\caption{ \label{sine}
Phase diagram for a sinusoidal pinning force $Y(x)=-\sin(x)$. The
IS-CM coexistence region is shaded gray. The $F=0$ region in which the
system can only exist in the CS phase is denoted by a series of
x's. The region of IS-CS coexistence is denoted by medium on-axis gray
shading. The IS$\rightarrow$CS and IS$\leftarrow$CS phase boundaries
are the points $(\mu=\mu_u=2, F=0)$ and $(\mu=\mu_d\approx 1.49, F=0)$
respectively. The line $F_{\uparrow}^i$ is the forces at which the 
system depins from the incoherent state. The line $F_{\downarrow}$ is 
the force at which a moving system stops upon lowering the drive.}
\end{figure}

Finally, for any continuous pinning force, the IM phase is not stable
even in the short time analysis.  (See Appendix D.) This result is
consistent with the findings of Strogatz and collaborators
\cite{Strogatz} as well as our simulations.

\section{Averaging over disorder}
\label{Averaging over disorder}

In this section we discuss the role of the shape of the distribution
$\rho(h)$ of pinning strengths on determining the nonequilibrium phase
diagram.  In the previous sections we restricted ourselves
to an infinitely sharp distribution, $\rho(h)=\delta(h-1)$.  This
choice is appropriate for systems with strong pinning and allows for a
direct comparison with the results of Strogatz and collaborators
\cite{Strogatz}.  It is easy to show that the nonequilibrium phase
diagram of the driven system retains the same qualitative structure
for any distribution that is sharply peaked around a finite value of
the pinning strength and vanishes below a finite $h_0>0$.  A broad
distribution of pinning strength may, however, qualitatively alter the
mean field physics. Broad distributions $\rho(h)$ are of interest to
model physical systems with weak pinning. Furthermore, a broad
distributions of pinning strengths yields variations of the local
stresses in the mean field theory and may give us some insight into
the behavior of the system in finite dimensions.

We consider a distribution of pinning strengths $\rho(h)$ that
vanishes below a minimum pinning strength $h_0\geq 0$. As will
become apparent below, it is important to distinguish three class of
distributions:
\begin{enumerate}
\item
distributions that vanish below a {\em finite} pinning strength, i.e.,
$\rho(h)=0$ for $h<h_0$, with $h_0>0$;
\item distributions with no finite lower bound of the pinning strength, 
but zero weight at $h=0$, i.e., $h_0=0$, and $\rho(0)=0$;
\item distributions with no finite lower bound of the pinning strength, 
and finite weight at $h=0$, i.e., $h_0=0$, but $\rho(0)>0$.
\end{enumerate}
The nonequilibrium phase diagram depends qualitatively on whether or
not the lower bound $h_0$ is finite. If the distribution of pinning
strengths $\rho(h)$ vanishes below a minimum pinning strength $h_0>0$,
the single particle depinning threshold $F_{\rm sp}$ remains finite
and the system exists in an IS phase for $F<F_{\rm sp}$. When $h_0=0$,
the single particle depinning threshold vanishes and the IS state can
only be stable at $F=0$.

If the IS phase exists, its stability can be analyzed for an arbitrary
distribution $\rho(h)$ by the perturbation theory described in
Sec. IV. For arbitrary $h$, the static force balance equation has the
form
\begin{equation}
\label{staticforce_h}
0=F-u\sin(\theta-\psi)+hY(\theta-\beta)\;,
\end{equation}
with the self-consistency condition given by
Eq.~(\ref{self_consistency1}).  Clearly this equation is identical to
the equation studied in Sec. IV for $h=1$, provided we rescale both
the driving force $F$ and the coupling strength $u$ by the pinning
strength $h$. We can then carry out the perturbation theory described
in Sec. IV as a perturbation theory in powers of $u/h$, provided of
course $u<<h_0$. This shows that the perturbation theory breaks down
when $h_0\rightarrow 0$. Furthermore we must require $F<F_{\rm sp}$,
which is a necessary condition for the existence of the IS
phase. Proceeding precisely as in Sec. IV and using the same notation,
we obtain an expression for the coherence $r$ as a power series in
$u/h$, given by
\begin{eqnarray} 
r=f(u,F)=\int dh \rho(h) [& & r_1(F/h)\frac{u}{h}+r_2(F/h)\Big(\frac{u}{h}\Big)^2+\nonumber\\ 
r_3(F/h)\Big(\frac{u}{h}\Big)^3+\dots]\;,
\label{f(u,F)_h} 
\end{eqnarray} 
with
\begin{subequations}
\begin{eqnarray}  
r_1(F/h)&=&{1\over2 {\tilde a}(F/h)}\;,\\ 
r_2(F/h)&=&0\;,\label{f(u)_2aF_r2}\\ 
r_3(F/h)&=&{3\over8}\frac{{\tilde a}(F/h){\tilde c}(F)-
2\big[{\tilde b}(F/h)\big]^2}{{\tilde a}(F/h)^5}\;,
\label{expanded_f(u)_F_h}
\end{eqnarray}
\end{subequations}
and 
\begin{subequations}
\begin{eqnarray}  
{\tilde a}(F/h)&=&Y'(\delta_0)\;,\\ 
{\tilde b}(F/h)&=&Y''(\delta_0)/2\;,\\
{\tilde c}(F/h)&=&Y'''(\delta_0)/6\;.
\end{eqnarray}
\label{expanded_Y(delta,F)_h}
\end{subequations}
The boundary of stability of the IS phase, $\mu_u(F)$,
is obtained like before by solving 
the implicit equation $r(\mu,F)=f(u=\mu r,F)$ with $f(u,F)$ given
by Eq.~(\ref{f(u,F)_h}), with the result
\begin{equation} 
\label{mu_u_broad}
\mu_u(F)=2\bigg[\int \rho(h) \frac{1}{h\tilde{a}(F/h)}\bigg]^{-1}\;.
\end{equation}
If the distribution $\rho(h)$ vanishes below a finite minimum
pinning force $h_0>0$, then $\mu_u$ 
remains finite and there is a range of 
$\mu$ and $F$ where the IS 
phase is stable. Conversely, if $h_0\rightarrow 0$, the integral in 
Eq.~(\ref{mu_u_broad}) may diverge, yielding $\mu_u=0$.
Below we will treat in detail the case of a piecewise linear 
pinning force, with $Y(x)=-ax$. In this case Eq.~(\ref{mu_u_broad}) reduces
to
\begin{equation}
\mu_u=2a\bigg[\int dh \frac{\rho(h)}{h}\bigg]^{-1}\;.
\label{mu_u_allrho_scallop}
\end{equation} 

For concreteness, we consider a distribution of the form 
\begin{eqnarray}
\label{broad_rho}
& & \rho(h)=(h-h_0)^\alpha\,e^{-(h-h_0)}\;,\hspace{0.2in}h\geq h_0\;,\nonumber\\
& & \rho(h)=0\;,\hspace{0.2in}h< h_0\;,
\end{eqnarray}
with $h_0>0$ and $\alpha>0$.  This form encompasses the three classes
of distribution functions introduced at the beginning of the
section. We can then obtain the boundary of the IS phase for a
piecewise linear pinning force by evaluating the integral on the right
hand side of Eq.~(\ref{mu_u_allrho_scallop}).  For distributions of
the first class, corresponding here to $h_0>0$ and $\alpha=0$, we find
that $\mu_u$ is finite at finite $F$ and it is given by $\mu_u=
2ae^{h_0}E_1(h_0)$, where $E_1(x)$ is the exponential integral.  For
this type of distribution it can be shown that the nonequilibrium
phase diagram remains qualitatively similar to the one obtained for
the sharply pinned distribution, $\rho(h)=\delta(h-1)$, even for all
types of pinning forces studied in Sec. V.  When $h_0\rightarrow 0$
the perturbation theory breaks down and the existence of a finite
value of $\mu_u$, even at $F=0$, depends on the form of $\rho(h)$ for
$h\rightarrow 0$.  For distributions of the second class, with
$h_0=0$, but $\rho(0)=0$, it can be shown that $\mu_u$ is finite at
$F=0$, but vanishes at all finite $F$. In this case there is an IS-CS
transition at $F=0$, which is a remnant of the transition seen at
finite $F$ for the case of an infinitely sharp pinning strength
distribution. For instance, for $\rho(h)=he^{-h} (\alpha=1)$, there is
an IS-CS transition at $F=0$ and $\mu_u\simeq 0.27$. Finally, for
distributions in the third class, with $\rho(0)>0$, it can be shown
that $\mu_u$ vanishes as $1/\ln(1/h_0)$ when $h_0\rightarrow 0$.  For
such distributions, there is no IS phase even at $F=0$.  The phase
diagrams for this class of distributions of pinning strength are
qualitatively different from those presented in Sec. V for all pinning
forces.  An example is shown in Fig.~\ref{exp} for the piecewise
linear pinning force and $\rho(h)=e^{-h}$.  This phase diagram has
been obtained numerically.  In the limit of large system sizes and
adiabatically slow ramp rates $dF/dt$, no IS phase is observed even at
$F=0$.  The small region of hysteresis in the transition between the
CS and CM phases is also washed out by the disorder averaging.  The
depinning curve $F^c_{\uparrow}$ displays a broad maximum at a finite
$\mu$ and vanishes as $\mu\rightarrow\infty$.
 
\begin{figure}
\begin{center}
\epsfxsize=8cm
\epsfbox{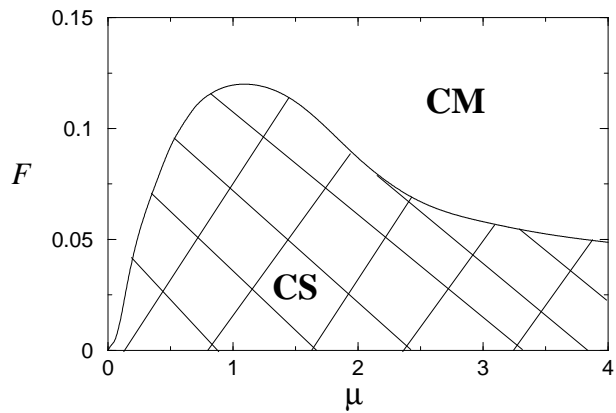}
\caption{\label{exp}
Phase diagram in the $\mu-F$ plane for a piecewise linear pinning force,
with $a=1/\pi$ and $c=0$, and $\rho(h)=
e^{-h}$. The depinning curve has been obtained numerically for a system with 
$N=1024$ and a ramp rate of $dF/dt=10^{-6}$.  }
\end{center}
\end{figure}

In general, the numerical simulations show that a broad distribution
of pinning strengths with vanishing $h_0$ always washes out the IS
phase and any hysteresis of the depinning transition. Whether this
behavior persists in finite dimensions remains an open question.

\section{Discussion}
\label{Conclusions}

In this paper we have used a combination of analytical and numerical
techniques to study the nonequilibrium mean field phase diagram of a
model of an extended systems with phase slips driven through disorder.
For uniform pinning, we generically find two stable static phases and
a single moving phase.  Both incoherent (IS) and coherent static (CS)
phases are possible, as well as regions where the two phases coexist.
The moving phase, in contrast, is always coherent (CM) in mean field
theory. (An incoherent moving phase can be prepared by using special
initial conditions, but does not appear to be stable.)  Coexistence of
two, or even three, of these phases can occur depending on the system
preparation; this coexistence results in hysteretic transitions.  Such
a variety of phases was not found for the case of a sinusoidal pinning
force analyzed earlier,\cite{Strogatz} where only the IS and CM phases
were found.  While a discontinuity in the pinning force is not
required for the existence of the new CS phase at large values of the
coupling constant $\mu$, a jump discontinuity in the pinning force
does increase the range of $F$ and $\mu$ over which the CS phase is
observed.  This is because discontinuity in the pinning force makes it
more difficult for the system to depin, so that the static pinned
phases can exist up to large coupling strengths, where the system is
forced to acquire long range coherence.  Once the system has become
coherent, and therefore more rigid, the depinning threshold decreases
with increasing $\mu$, but remains finite for all finite values of the
coupling strength and only vanishes for $\mu\rightarrow\infty$.  For a
continuous pinning forces, on the other hand, the depinning threshold
vanishes above a finite value of $\mu$.

In order to make some contact with particle simulations and with
experiments, it is useful to discuss the mean field phase diagram in
terms of the disorder strength $h$ and the driving force $F$, rather
than in the $(\mu,F)$ plane as done so far.  In most particle
simulations it is the strength of the disorder that is most easily
varied rather than the strength of the coupling. Disorder is also a
crucial control parameter in many experimental systems. For
instance,varying the applied magnetic field in current-driven vortex
lattices has the effect of varying the strength of the disorder. At
high fields the vortex lattice becomes softer and can better adjust to
disorder. Increasing the magnetic field therefore corresponds to an
effective increase of the disorder strength.  Fig.~\ref{fhphasedia}
shows the mean field phase diagram in the $(h,F)$ plane for the
discontinuous soft cubic pinning force shown in Fig. 1b. The
corresponding phase diagram in the $(\mu,F)$ plane was shown in
Fig. 2.

\begin{figure}
\begin{center}
\epsfxsize=8cm
\epsfbox{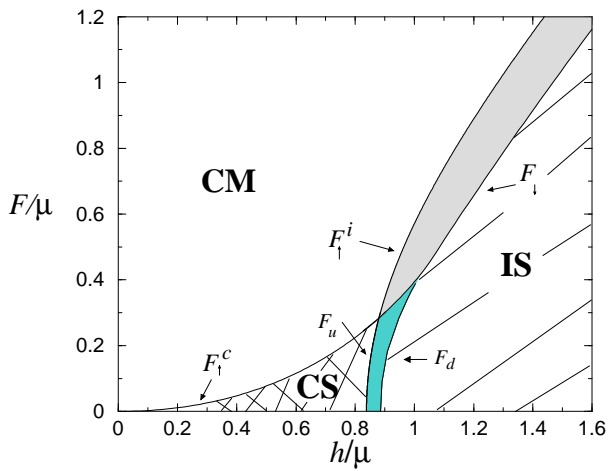}
\end{center}
\caption{ \label{fhphasedia}
Phase diagram, redrawn in the disorder-drive plane,
for a discontinuous soft cubic
pinning force of the type shown in Fig. 1b and
$\rho(h)=\delta(h-1)$. The disorder $h$ and drive $F$ are normalized
by the strength of the phase-slip interaction, $\mu$.
The parameter values and the symbols are the
same as in Fig. 2.}
\end{figure}

When the disorder is weak relative to the strength of the coupling
$\mu$ the static phase is coherent. At strong disorder the static
phase is incoherent.  The transition between the coherent and
incoherent static phases at fixed $\mu$ is hysteretic with a region of
coexistence of the two phases.  At weak disorder there is a continuous
``elastic-like'' depinning transition from the CS to the CM phase. At
large disorder the static phase is incoherent and degrees of freedom
depin independently at the single particle depinning threshold,
$F^i_\uparrow$. The moving system immediately acquires long-range
correlations, becoming much stiffer and harder to pin. As a result,
when the force is ramped down the CM state repins at the lower force
$F_\downarrow$.  The qualitative features of this phase diagram are
remarkably similar to those obtained by Olson and collaborators
\cite{Olson01} in a numerical simulation of a model of a
current-driven layered superconductors, with magnetically interacting
pancake vortices. At weak disorder these authors find that the layers
are coupled and the system forms a coherent three-dimensional static
phase, with long-range correlations along the direction normal to the
layers, which depins continuously. At strong disorder the static state
consists of decoupled two-dimensional layers. When the driving force
is ramped up from this incoherent static state, the layers depin
independently at the single-layer depinning threshold and the
transition is hysteretic. One difference between our mean field model
and the numerical model studied by Olson {\it et al.} is the absence, in our
model, of an incoherent moving phase. In the layered superconductor at
strong disorder the layers remain decoupled upon depinning up to a
second, higher threshold force where a dynamical re-coupling
transition occurs. Finally, these authors also observe a sharp
increase in the depinning threshold at the crossover or transition
from continuous to hysteretic depinning, not unlike that shown in
Fig.~\ref{fhphasedia}.  A strong crossover from elastic to plastic
with increasing disorder strength, with an associated sharp rise of
the depinning threshold, has also been seen in a variety of two
dimensional simulations, such as those by Faleski, {\it et
al.}\cite{FMM} Macroscopic hysteresis has not, however, been observed
in these two-dimensional models. Our work suggests that mean field
models with strong disorder tend to overestimate hysteresis. In mean
field there is no range of correlation lengths and hysteresis will
always occur when the system is driven from a strongly pinned
incoherent phase, where all degrees of freedom depin independently at
the single particle depinning threshold. Upon depinning, the system
acquires long range order and becomes therefore much stiffer, so that
when the force is ramped down it can remain in the sliding state down
to much lower values of the driving force.
  
Early transport experiments on current-driven vortices in NbSe$_3$
showed S-shaped IV characteristics at high magnetic fields with a
peak in the differential resistance as a function of driving
current.\cite{fingerprint} Other puzzling effects were observed in the
region of the peak, including unusual frequency dependence of the ac
response and fingerprint phenomena.  These experimental findings were
originally interpreted in terms of plastic depinning of the vortex
system and macroscopic coexistence of disordered and ordered bulk
vortex phases.  This interpretation was corroborated by a number of
simulations in two-dimensions, where the crossover from elastic to
plastic depinning is clearly seen as a function of disorder
strength. For strong disorder the system exists in a disorder static
phase that depins plastically and then undergoes a dynamical ordering
transition to a moving ordered phase. The peak in the differential
resistance corresponds to such a dynamical ordering transition and in
simulations is clearly associated with a sharp drop in the number of
topological defects in the driven lattice.  More recent experiments
have suggested, however, that the disordered phase is a metastable
phase that is injected at the sample's edges and then anneals into the
stable elastic phase as it gets driven into the
sample. \cite{paltiel,Paltiel02,marchevsky} This interpretation has
been confirmed by comparing transport experiments in the conventional
strip geometry, where the edge effect is always present, to
experiments in a Corbino disk geometry, where the vortices are driven
to move in concentric circular orbits in a disk-shaped sample,
eliminating boundary effects.  Although there is mounting experimental
evidence that these edge contamination effects may indeed control much
of the vortex dynamics observed in experiments, the comparison with
simulations, where coexistence of bulk ordered and disordered phases
is routinely observed, remains puzzling.  Of course one important
difference is that most of the simulations are carried out at zero or
very low temperature, where the disordered phase may be artificially
stabilized.

Substantial phase slip effects have also been observed in CDW systems,
especially at the contacts \cite{Thorneexpt}, and have been associated
with the ``switching'' observed in certain materials.  The reported
correlation between broadband noise and macroscopic velocity
inhomogeneities also supports the idea that in these systems the
dynamics may be dominated by large scale plasticity.  \cite{broadband}
While the switching itself has also been explained as arising from the
presence of normal carriers \cite{levy92}, phase slips seem crucial to
account for the correlation between broadband noise and macroscopic
velocity inhomogeneities.
 
Finally, similar behavior has also been observed in colloids driven
over a disordered substrate.  Pertsinidis and Ling \cite{Ling} have
studied experimentally single layers of two-dimensional colloid
crystal driven by an electric field over a disordered substrate.  They
observe plastic-like or filamentary flow of the colloids, with a
velocity-force curve that is always convex upward and shows no
hysteresis.  Langevin simulations by Reichhardt and Olson
\cite{ReichhardtOlson} find a sharp crossover from elastic to plastic
depinning as the strength of substrate is increased.  Though the
direct applicability of our mean field model and results to
experimental systems remains to be proven, this work lays out a detailed
foundation for understanding the role of phase slips and
topological defects on the dynamics of driven disordered systems.
Preliminary numerical studies of the phase slip model in three
dimensions, with a sinusoidal pinning potential, suggest that the
depinning transition may not be hysteretic in the thermodynamic limit.
This is similar to what suggested by studying the mean field with a
broad distribution of pinning strengths, as shown in Fig.~\ref{exp},
where the distribution of pinning forces the incoherent static (IS)
phase.  Clearly more work is needed to establish if such a finding is
generic in finite dimensions.  An important open question is whether
the transition from elastic to plastic depinning (with or without
macroscopic hysteresis) is a crossover or is associated with some type
of tricritical point, as suggested by the present and other mean field
models.

This work was supported in part by NSF grants DMR-9730678, DMR-0109164 and 
DMR-0305407.

\appendix

\section{Coherence at $F=0$}
\label{calcA}

In this appendix we describe the calculation of the coherence $r(\mu)$
of static states at $F=0$. First we derive an expression for the
function $f(u)$ defined in Eq.~({\ref{self_consistency1a}) for an
arbitrary pinning force, $Y(\delta)$.  Once $f(u)$ is known, the
coherence is then obtained by solving the self-consistency condition,
$r=f(\mu r)$. The calculation is complicated by multivalued solutions to
the self-consistency equations, which leads to multiple metastable
states. A consistent selection principle is applied,
namely, choosing the coherence $u$ to be maximal, given $\mu$.
The range of available metastable states is also used
to determine $\mu_T$, the value of coupling above which the depinning
field is zero.

\subsection{Change of variables}

As discussed in Section \ref{COV}, it is 
convenient to perform a change of variables in Eq.~({\ref{self_consistency1a})
and integrate over  $\delta$ rather than over the random phase $\beta$.
The function $f(u)$ is then given by
\begin{eqnarray}
f(u)={1\over2\pi}\int_{-\pi}^{\pi} d \delta\left({\partial \beta \over \partial \delta}\right) \cos[\delta+\beta(\delta,u)]\;,
\label{r(u)_1a}
\end{eqnarray}
where $u=\mu r$.  Since $Y(\delta)$ 
is $2\pi$ periodic, the integration in Eq.~(\ref{r(u)_1a}) 
can be carried out over
any $2\pi$ interval. Here we choose the interval $[-\pi,\pi]$.
The change of variable allows us to evaluate $f(u)$ analytically 
as the force balance equation,
Eq.~(\ref{force_balance1}), while transcendental in $\delta(\beta,u)$,
is simply a linear equation in the phase $\beta(\delta,u)$.
We can therefore immediately write the solution 
$\beta(\delta,u)$ of Eq.~(\ref{force_balance1}),
substitute it in Eq.~(\ref{r(u)_1a}), and evaluate the integral 
to obtain $f(u)$.
As we will see below, the only difficulty in carrying out this program 
is that the 
phase $\delta(\beta,u)$ is generally a multivalued function of $\beta$.
Therefore care must be taken in selecting the portion of the curve that
must be included in the integral. The choice  is dictated by the requirement
that the imaginary part of the self consistency condition, which now reads
\begin{eqnarray}
0={1\over2\pi}\int_{-\pi}^{\pi} d \delta\left({\partial \beta \over \partial \delta}\right) \sin[\delta+\beta(\delta,u)]\;,
\label{imag_1a}
\end{eqnarray}
be satisfied, and that the phase $\delta(\beta)$ span a full $2\pi$ 
interval in $\beta$. 

For static solutions and $F=0$ the balance equation (\ref{force_balance1})
can be written as
\begin{equation}
\sin(\delta+\beta)=\frac{Y(\delta)}{u}\;.
\label{balance_eqn_appA}
\end{equation}
Since $|\sin(\delta+\beta)|\leq 1$, the right hand side of 
Eq.~(\ref{balance_eqn_appA}) must also be bounded in magnitude by one.
This means that all solutions to Eq.~(\ref{balance_eqn_appA})
must satisfy
\begin{equation} 
-\delta_{max}(u)\leq \delta \leq \delta_{max}(u)\;,
\end{equation}
where $\delta_{max}(u)$ is
defined by
\begin{equation}
\label{delta_max(u)}
|Y(\delta_{\max})|=u\;,
\end{equation}
or $\delta_{max}(u)\equiv|Y^{-1}(u)|$, 
with $Y^{-1}$ denoting the inverse function.
Note that if $Y(\delta)$ is non-monotonic in the
interval $[-\pi,\pi]$, as it is for instance the case for the soft
cubic potential shown in Fig.~\ref{potentials}(b), 
then for $u>|Y(\pi)|$ there are two possible
values of $|Y^{-1}(u)|$ in the range $[0,\pi]$. In this case 
$\delta_{max}$ is defined as the smallest of these two values. 
At the end of this
Appendix we will discuss the relevance of the second solution and
demonstrate that it corresponds an unstable state.

\subsection{Metastable states}

For every fixed value of $u$, there is in general an infinite set of 
solutions for the phase $\delta$ in the range $[-\delta_{max},\delta_{max}]$. 
The corresponding solutions for the phase $\beta$ as a function of $\delta$
can be enumerated by indexing them with an integer, $n$. They are given by
\begin{eqnarray}
\beta_n(\delta,u)= -\delta + n \pi + (-1)^n\sin^{-1}(Y(\delta)/u) \;,
\label{delta(beta)}
\end{eqnarray}
where we only consider values of the function $\sin^{-1}(x)$ in the
range $[-\pi/2,\pi/2]$. Since the calculation of
$\beta_n(\delta,u)$ and $\delta(\beta,u)$ is carried out at fixed $u$, 
from here on 
we will simply omit the $u$ dependence in the argument of these functions.
The typical behavior of the phase $\delta$ as a function of
$\beta_n$, for $-2\leq n \leq 2$, is shown in Fig.~\ref{beta)delta_fig}.  

The integral in Eq.(\ref{r(u)_1a}) must span a full period (in $\beta$) of the 
$\delta(\beta)$ curve. As evident from Fig.~\ref{beta)delta_fig}, 
this always corresponds to a pair of consecutive
even-odd sections.   
Here we choose to work with the $n=0$
section, and the upper and lower halves of the $n=-1$ and the $n=1$
sections, respectively.  
This choice is
equivalent, for instance, to that of the $n=0$ and the full $n=-1$ 
sections (or $n=0$ and $n=1$),
but it has the advantage of being symmetric about the origin.
The chosen portion of the $\delta(\beta)$ curve is displayed in 
Fig.~\ref{F=0_delta(beta)} 
for three different values of
$u$. The figure shows how the phase becomes multivalued as $u$ is increased.

\begin{figure}
\begin{center} 
\epsfxsize=8cm
\epsfbox{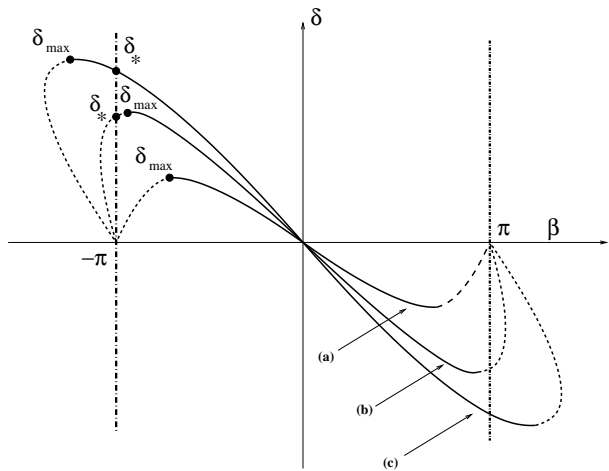}
\caption{The figure shows the behavior of the phase $\delta(\beta)$ 
for three values of $u$. The $n=\pm 1$ half-sections are dashed, 
while the $n=0$ 
section is solid. Curve (a) corresponds to $u\leq a$ and is single valued. 
Curves (b) and (c) are both multivalued and correspond to (b)
$a<u\leq Y(\pi/2)$ and (c) $u> Y(\pi/2)$. 
The section $\beta_0$ ends at the points $\pm\delta_{max}$, where the 
half sections $\beta=\mp 1$ begin.
For curve (b) these points lie within  the portion of the curve that must be
included in the integral to determine $f(u)$. For curve (c) they lie outside. 
$\delta_*$ denotes the non-zero value of the phase at $\beta=-\pi$.}
\label{F=0_delta(beta)}  
\end{center}
\end{figure} 

For $u\leq a$, 
with $a$ the linear slope of the pinning force $Y(\delta)$
at $\delta=0$, the phase $\delta$ is single-valued, as in curve (a)
of Fig.~\ref{F=0_delta(beta)}. In this case
integrating over a full period in $\beta$ is equivalent to
integrating over the entire curve, consisting of the full $n=0$ central 
section (solid) and the two $n=\pm 1$ half sections (dashed). Making use of 
the symmetry of the integrand about $\delta=0$, we obtain
\begin{eqnarray}
f(u)&=&{1\over\pi}\int_0^{\delta_{max}} d \delta\left({d \beta_{-1} \over d \delta}\right) \cos[\delta+\beta_{-1}(\delta,u)]\nonumber\\
&+&{1\over\pi}\int_{\delta_{max}}^0 d \delta \left({d \beta_0 \over d \delta}\right) \cos[\delta+\beta_{0}(\delta,u)]\;.
\label{r(u)_2}
\end{eqnarray}
Upon substituting the expressions for 
$\beta_{-1}(\delta)$ and $\beta_0(\delta)$ from Eq.~(\ref{delta(beta)})
in Eq.~(\ref{r(u)_2}),
we obtain
\begin{eqnarray}
f(u)={2\over\pi}\int_0^{\delta_{max}} d \delta~\sqrt{1-[Y(\delta)/u]^2}\;, \hspace{0.25in} u\leq a  \;.
\label{r(u)_3}
\end{eqnarray}
When $\delta(\beta)$ is single valued, the integral in 
Eq.~(\ref{self_consistency1b}) over the entire period gives zero, 
so that the imaginary part of the self consistency condition is 
satisfied.

When $u \geq a$, the phase $\delta$ is multivalued, as exemplified in cases (b)
and (c) in Fig.~\ref{F=0_delta(beta)}. In this case
one can no longer simply integrate over the full curve in the 
range $\beta\in [-\pi,\pi]$.
Rather, one must select a portion, of measure
$2\pi$ in $\beta$, that satisfies the imaginary part of the self consistency
condition, Eq.~(\ref{imag_1a}). 
As discussed in Sec.~\ref{COV} we choose the portion
of the curve corresponding to the metastable states 
that would be accessed by
adiabatically increasing $u$ from zero. For $\psi=0$, this choice
corresponds to the connected part of the $\delta(\beta)$ curve lying
between $\beta=-\pi$ and $\pi$. This choice is odd about the
origin and therefore automatically 
satisfies  Eq.~(\ref{imag_1a}). 
The phase $\delta$ now has two values at $\beta=\pi$, 
$\delta=0$ and $\delta=\delta_*$, which is defined implicitly 
as the nonzero root of the equation
\begin{eqnarray}
-Y(\delta_*)=u\sin(\delta_*)\;.
 \label{delta_star}
\end{eqnarray}
The value $\delta_*$ is the desired upper limit in the integration 
over $\delta$ in Eq.~(\ref{r(u)_1a}).
When $a \leq u \leq Y(\pi/2)$, corresponding to curve (b) in
Fig.~\ref{F=0_delta(beta)}, the root $\delta_*$ is smaller than $\delta_{max}$
and the portion of the curve to be included in the integrand spans the entire
$\beta_0(\delta)$ section (solid line) {\em and} those parts of
the $\beta_{\pm 1}(\delta)$ half sections (dashed) that lie within
$\beta=[-\pi,\pi]$. For this range of $u$ values
we find
\begin{eqnarray}
f(u)&=&{2\over\pi}\int_0^{\delta_{max}} d \delta~\sqrt{1-(Y(\delta)/u)^2}-{Y(\delta_*)\over u\pi} \nonumber\\
&-&{1\over\pi}\int_0^{\delta_*} d \delta~\sqrt{1-(Y(\delta)/u)^2}\;,\nonumber\\
& & \hspace{1.in}  a\leq u\leq |Y(\pi/2)| \;.
\label{r(u)_4}
\end{eqnarray}

At $u=|Y(\pi/2)|$, $\delta_*=\delta_{max}$. For $u >
|Y(\pi/2)|$, corresponding to the situation illustrated in curve (c), 
$\delta_{max}$ exceeds $\delta_*$ and 
the portion of the curve to be included in the integrand 
only spans that part of
the $\beta_0(\delta)$ section (solid) that lies in $\beta\in[-\pi,\pi]$, as 
seen from
Fig.~\ref{F=0_delta(beta)}. In this case we obtain
\begin{eqnarray}
f(u)&=&{1\over\pi}\int_0^{\delta_*} d \delta~\sqrt{1-(Y(\delta)/u)^2}
-{Y(\delta_*)\over u\pi} \;,\nonumber\\
& &  \hspace*{0.8in} u\geq |Y(\pi/2)| \;.
\label{r(u)_5}
\end{eqnarray}
The three equations, Eqs.~(\ref{r(u)_3}), (\ref{r(u)_4}) and
(\ref{r(u)_5}), give the function $f(u)$ at all $u$ for an arbitrary pinning
force, $Y(\delta)$. It can be shown that when Eq.~(\ref{r(u)_3}) is
expanded for small $u$, the perturbative result,
Eq.~(\ref{expanded_f(u)}), is recovered.

For a piecewise linear pinning force, with $Y(\delta) = -a \delta$ for $-\pi \leq \delta \leq \pi$, the integrals in Eqs.~(\ref{r(u)_3}), (\ref{r(u)_4}) and (\ref{r(u)_5}) can be evaluated analytically, with the result
\begin{equation}
f(u)=
\begin{cases}
{u \over 2a}, &
            u\leq a, \\
{u \over 2a} +{\delta_*\over 2\pi}
           \left({2a\over u} - {u\over a}- \cos \delta_* \right), & 
            a < u \leq a\pi/2,\\
{\delta_*\over 2\pi}
           \left({2a\over u} + {u\over a} + \cos \delta_* \right), &
           u < a\pi/2,
\end{cases}
\label{sawtooth_Y_r(u)}
\end{equation}
where $\delta_*=(u/a)\sin(\delta_*)$. The coherence $r$ is then
determined by the solution of $r=f(\mu r)$. For $u\leq a$
the equation for the coherence is $r=\mu r/\mu_u$, where $\mu_u=1/(2a)$.
If $\mu \neq \mu_u$, the only solution is $r=0$.
For $\mu=\mu_u$ the equation is satisfied by any nonzero value of $r$ 
consistent with
$u\leq a$, or equivalently $r\leq 1/2$. Thus, at  $\mu=\mu_u$
the coherence jumps discontinuously from zero to the value $r_0=1/2$.  
By expanding $f(u)$ for  $u\rightarrow a^+$ 
we find
that for $\mu \agt \mu_u$,
\begin{equation} 
r-r_0\propto(\mu-\mu_u)^{2\over5}\;.
\label{r_near_mu_0} 
\end{equation} 
The full solution $r$ as a function of $\mu$ is shown in Fig.~\ref{r(mu)}.

We now return to the question of the existence of solutions $\delta(\beta)$
outside the range $[-\delta_{max},\delta_{max}]$. This is relevant for
pinning forces $Y(\delta)$
that are non-monotonic in the interval $[-\pi,\pi]$. For such pinning forces
the Eq.~(\ref{delta_star}) has two nonvanishing solutions. 
The smallest of these
two solutions, $\delta_*$ defines the range of phases that have
been used in the calculation of the coherence described above. 
Denoting the largest of the two solutions by $\delta_u$, we note that for 
$u>|Y(\pi)|$ there will also be solutions for the phase $\delta$ 
lying in the ranges
$[\delta_u,\pi]$ and $[-\pi,-\delta_u]$. Examples of such solutions are
shown in Fig.~\ref{unstable_delta} for the soft cubic pinning force.
The solutions outside the range $-\delta_{max}\leq\delta\leq\delta_{max}$
are the top and bottom branches in the figure. It can be shown
that such solutions are always unstable, while the center branch is stable.
This is easily seen by plotting the total force 
$F_{tot}=-u \sin (\delta+\beta) +Y(\delta)$ acting on a
domain versus the phase $\delta$, for a fixed value $\beta$. 
The stable solutions of the force balance equation  are the zeros 
of $F_{tot}(\delta)$ with a negative slope, so that they correspond to
minima of the total potential. The zeros with a positive slope are 
maxima of the potential and therefore represent unstable solutions. 
Of the two zeros shown for instance in Fig.~\ref{totforce} for $\beta=\pi/2$,
only the left solution, which lies in the range $[-\delta_{max},\delta_{max}]$ 
is stable, while the right one is outside this range and is unstable. 
Changing the value of $\beta$ would simply shift the curve of $F_{tot}$
along the $\delta$ axis, with the stable root always remaining inside 
the interval $[-\delta_{max},\delta_{max}]$.

\begin{figure}
\begin{center}
\epsfxsize=8cm
\epsfbox{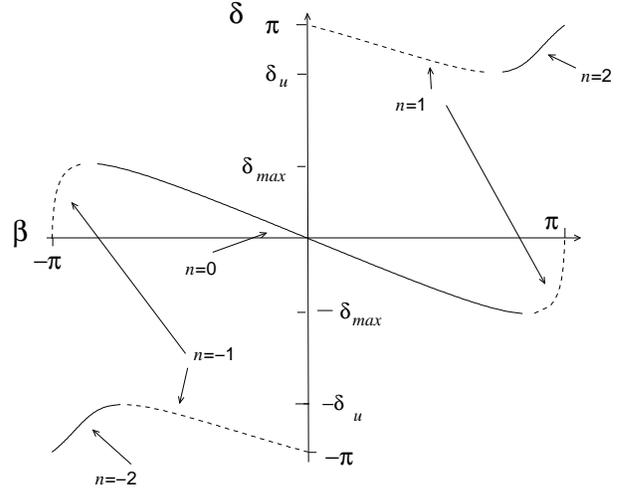} 
\caption{The figure shows the phase $\delta$ versus $\beta_n$ at $u=0.8$, 
for $n=0,\pm 1, \pm 2$, for the continuous pinning
force of Fig.~\ref{potentials}(c), with $a=3\sqrt{3}/(2\pi)$ and
$c=-3\sqrt{3}/(2\pi^3)$. The upper and lower branches, lying outside 
the range $[-\delta_{max},\delta_{max}]$ are unstable, while the 
central branch is stable.}
\label{unstable_delta}  
\end{center}
\end{figure} 

\begin{figure}
\begin{center} 
\epsfxsize=8cm
\epsfbox{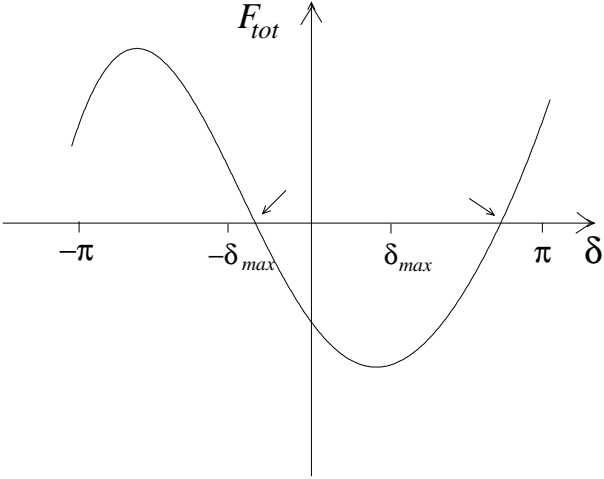} 
\end{center}
\caption{
Plot of $F_{\bf tot}$ versus $\delta$ for $\beta=\pi/2$, $u=0.8$
for the continuous pinning force of Fig.~\ref{potentials}(c), 
with $a=3\sqrt{3}/(2\pi)$ and $c=-3\sqrt{3}/(2\pi^3)$. 
The equation $F_{\bf tot}=0$ has two solutions. The left solution, with 
negative slope is 
stable, while the right one, with positive slope, is unstable.}
\label{totforce}  
\end{figure} 

\subsection{Derivation of $\mu_T$}
The number of metastable static states available to the system
plays an important role in determining the depinning threshold.
In general the system can exist in a large number of static metastable states
and the function $\delta(\beta)$ becomes more multivalued as $u$ increases,
as shown in Fig.~\ref{F=0_delta(beta)}.
The number of metastable states is not, however, a monotonic function
of $u$ as only values of $\delta$ lying in the interval
$[-\pi,\pi]$ are acceptable solutions.
The number of available metastable
states increases with increasing $u$ until $\delta_{max}(u)=\pi$,
corresponding to $u=|Y(\pi)|$. 
As $u$ is increased beyond $|Y(\pi)|$ 
the number of metastable states decreases.  When an infinitesimal
force is applied, all the phases are pushed forward and an infinitesimal 
number of static metastable states becomes unstable 
as they can no longer satisfy the 
self-consistency condition. The system remains, however, pinned, provided 
there exist other static states that are still metastable. When $\delta_*(u)=\pi$, the situation changes as
there is {\em only one} 
metastable static solution that becomes unstable as soon
as an infinitesimal driving force is applied to the system. 
The system depins as soon as $F>0$, i.e., the threshold force
for depinning is zero.

It can be seen from Eq.~(\ref{delta_star}) defining $\delta_*$ that
for pinning forces
with $|Y(\pi)|>0$, $\delta_*<\pi$ for any finite
$u$. In this case $\delta_*$ approaches $\pi$ only in the limit
$u\rightarrow \infty$. Since $r$ is always finite, it is only in the
limit of infinite $\mu$ that the system approaches a perfectly ordered
floating state and the depinning threshold force goes to zero.
For continuous pinning forces with $Y(\pi)=0$,
$\delta_*=\pi$ at a {\em finite} value of $u=u_T\equiv
Y^{\prime}(\pi)$. For $u\geq u_T$, the system has
only a single, albeit partially disordered, state available. This state 
becomes unstable upon application of an infinitesimal driving force, and 
the system begins to slide. In other words, the threshold for depinning 
vanishes for all $u\geq u_T$ or, equivalently, all  
$\mu\geq\mu_T=u_T/f(u_T)$. Using Eq.~(\ref{r(u)_5})
we find the value of $\mu_T$ displayed in Eq.~(\ref{mu_T}).

\section{Depinning force $F_\uparrow^c(\mu)$}
\label{calcB}

In this appendix we calculate the depinning force 
$F_\uparrow^c(\mu)$ for hard and soft
cubic pinning forces, of the type sketched in Fig.~\ref{potentials}.
These forces are given by Eq.~(\ref{Y(x)}) with
$c>0$ for the hard cubic force and $c<0$ for the soft cubic
force. Due to the periodicity of the problem, we can 
restrict ourselves to any interval of $\delta$ of range $2\pi$.
For simplicity we choose again $\delta$ to lie in the $[-\pi,\pi]$ interval. 
In this interval the
force balance equation, with $\psi=0$ is
\begin{equation} 
0= F - u \sin (\delta+\beta) + Y(\delta)\;, 
\label{force_balance_F_n=0} 
\end{equation} 
and only
solutions to Eq.~(\ref{force_balance_F_n=0}) which
satisfy $-\pi \leq \delta(\beta,u,F) \leq \pi$ should be considered. 

 As
for the case $F=0$, the transcendental nature of the force balance equation,
Eq.~(\ref{force_balance_F_n=0}), can be circumvented by  integrating over
$\delta$ rather than over the phase $\beta$ in the self-consistency
conditions. Solving for $\beta(\delta,u,F)$ gives an infinite set of
of solutions, labeled by an integer $n$,
\begin{eqnarray}
\beta_n(\delta)= -\delta + n \pi + (-1)^n\sin^{-1}\left({Y(\delta)-F\over u}\right) \;,
\label{Fdelta(beta)}
\end{eqnarray}
where $\delta$ is restricted to lie in the range
\begin{equation}
\label{deltaF_range}
\delta_{min}(u,F)\leq\delta\leq\delta_{max}(u,F)\;,
\end{equation}
with 
\begin{eqnarray}
\label{deltamin_deltamax}
& & \delta_{min}(u,F)\equiv -Y^{-1}(F-u)\;,\nonumber\\
& & \delta_{max}(u,F)\equiv -Y^{-1}(F+u)\;.
\end{eqnarray}
The solution must satisfy the real and imaginary parts
of the self consistency condition, given by,
\begin{eqnarray}
& &r=f(u,F)\;,
\label{Real_appB}\\
& &0={1\over2\pi}\int_{2\pi} 
   d\delta\left({\partial \beta \over \partial \delta}\right) 
    \sin[\delta+\beta(\delta,u,F)]\;,
\label{Imag_appB}
\end{eqnarray}
with
\begin{equation}
f(u,F)={1\over2\pi}\int_{2\pi} 
  d\delta\left({\partial \beta \over \partial \delta}\right) 
   \cos[\delta+\beta(\delta,u,F)]\;.
\label{f(u,F)_appB}
\end{equation}
Throughout the analysis we will be
considering $\delta(\beta,u,F)$ for fixed values of $u$ and $F$. We
will therefore write $\delta=\delta(\beta)$, with the dependence on $u$ and $F$
implied.

As in the case $F=0$, the phase $\delta$ is generally a multivalued function
of $\beta$ (see Fig.~\ref{F_delta(beta)}). We consider only the metastable 
state corresponding to a connected portion of the curve 
$\delta(\beta)$ in the range $\delta\in[\delta_L,\delta_R]$, and it is 
this portion that is integrated over in the self-consistency conditions. We focus in this particular state because it is the one that controls depinning.
The points $\delta_L$ and $\delta_R$
bounding this portion are functions of $u$ and $F$ and may in general differ
from $\delta_{min}$ and $\delta_{max}$. They are determined by the requirement
that the imaginary part of the self-consistency condition, 
Eq.~(\ref{Imag_appB}), be satisfied and by
the condition that the portion of the function $\delta(\beta)$
bounded by these points span a full $2\pi$ interval
in $\beta$, i.e.,
\begin{equation}
\beta(\delta_L)+2\pi=\beta(\delta_R)\;.
\end{equation}
The details on how the limits of integration are determined 
and the corresponding
portion of the solution for $\beta(\delta)$ is chosen in each case are
given below.

\begin{figure} 
\begin{center}
\epsfxsize=8cm
\epsfbox{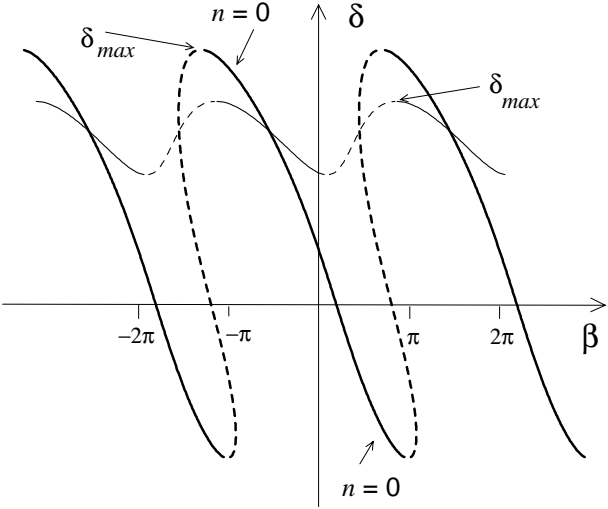}
\caption{The phase $\delta(\beta)$ as a function of  $\beta_n$ for 
$-2 \leq n \leq 2$ for two sets of values of $(u,F)$, corresponding to
single-valued (for $(u=0.1,F=0.2)$) and multi-valued (for $(u=0.35,F=0.2)$) 
solutions. Even $n$ branches are drawn 
as solid lines and odd $n$ branches are dashed.}
\label{F_delta(beta)}  
\end{center}
\end{figure} 

After evaluating the coherence, we can then proceed to compute the 
force $F_\uparrow^c(\mu)$ where the static coherent state becomes unstable
and the system begins to slide. As $F$ is increased at fixed $u$, the whole 
$\delta(\beta)$ curve shifts upward and both $\delta_{max}$ and $\delta_L$ 
increase. The number of static metastable states
in the range $\delta\in[-\pi,\pi]$ decreases, until at the critical force 
$F_\uparrow^c$ only one static metastable state remains. This occurs where the 
largest value of $\delta$ on the connected portion, denoted by $\delta_u$, 
reaches $\pi$, i.e.,  
\begin{equation}
\label{Fup_defined} 
\delta_u(u,F_\uparrow^c)=\pi\;.
\end{equation}
Upon further increasing $F$ the system depins. Equation~(\ref{Fup_defined}) 
defines
the boundary of stability of the coherent static state,
i.e., the depinning threshold, and can be solved to obtain $F_\uparrow^c(u)$. 
It will be shown below that, depending of 
the value of $u$, the connected portion satisfying the self consistency 
condition may or may not include $\delta_{max}$. For small values of $u$ 
it will and $\delta_u=\delta_{max}$. At larger values of $u$, 
the connected piece does not include $\delta_{max}$ and 
$\delta_u=\delta_L$.
Finally, the depinning threshold $F_\uparrow^c(\mu)$ as a function of $\mu$
is obtained by eliminating $u$ between the equation for the coherence
at threshold, $r=f(u,F_\uparrow^c)$ and the expression for $F_\uparrow^c(u)$
obtained from Eq.~(\ref{Fup_defined}).

\subsection{$F_\uparrow^c(\mu)$ for monotonic $Y(\delta)$}

The monotonic class consists of all hard cubic pinning forces and
those soft cubic pinning forces which have $|c| \geq a/\pi^{2/3}$.
Since the function
$Y(\delta)$ is monotonic, its inverse, $Y^{-1}(x)$, is single-valued
in  the entire range of
interest,  $-1 \leq x \leq 1$. 

A full period of $\delta(\beta)$ corresponds to a pair of consecutive
even-odd sections in $n$. In Fig.(\ref{F_delta(beta)}) we show plots
(with even sections shown as solid lines and odd sections shown dashed) 
of $\delta$ versus
$\beta_n(\delta)$ for two pair of values $(u,F)$, chosen so that in one case
the solution is single-valued and in the other it is multi-valued. 
In both cases the curves lack the symmetry of those for
$F=0$.
In general, the value of $u$ at which
$\delta(\beta)$ becomes multivalued depends on $F$.
At this value, denoted by $u_{sv}(F)$,  each
odd $\beta_n(\delta)$ develops an inflexion point
at $\delta=\delta_e$. In particular, for $n=1$, this requires
\begin{subequations}
\begin{eqnarray}
\left({\partial \beta_1(\delta,u_{sv},F) \over \partial \delta}\right)_{\delta=\delta_e}&=&0\;,\label{u_sv1a}\\
\left({\partial^2 \beta_1(\delta,u_{sv},F) \over \partial \delta^2}\right)_{\delta=\delta_e}&=&0\;.\label{u_sv1b} 
\end{eqnarray}
\end{subequations}
Using Eq.~(\ref{Fdelta(beta)}) for $\beta_1(\delta,u,F)$ we obtain the 
following pair of equations,
\begin{subequations}
\begin{eqnarray}
(u_{sv})^2&=&\big[Y'(\delta_e)\big]^2+\big[Y''(\delta_e)\big]^2\;,\label{u_sv2}\\
F&=&-Y(\delta_e)-Y''(\delta_e)\;,\label{u_sv1} 
\end{eqnarray}
\end{subequations}
which can be  solved to determine $u_{sv}(F)$. 

For $u<u_{sv}(F)$ the function $\delta(\beta)$ is single valued,
as shown in Fig.~\ref{delta(beta,u,F)_fourplots}(a).
Integrating over a $2\pi$ interval of
$\beta$ is equivalent to integrating over a full odd and even section.
We choose $\delta_L=\delta_{max}$, which requires
$\delta_R=\delta_{max}$ and automatically satisfies
the imaginary part of the self consistency condition.
The function $f(u,F)$ is then given by
\begin{eqnarray}
f(u,F)&=&
  \int_{\delta_{max}}^{\delta_{min}} 
    d \delta\left({\partial \beta_0 \over \partial \delta}\right) 
      \cos[\delta+\beta_0(\delta)]\nonumber\\
&+&\int_{\delta_{min}}^{\delta_{max}} 
    d \delta\left({\partial \beta_1 \over \partial \delta}\right) 
      \cos[\delta+\beta_1(\delta)]\;.
\label{r(u)_1*}
\end{eqnarray}
Equation~(\ref{r(u)_1*}) can be simplified as
\begin{eqnarray}
f(u,F)&=&{1\over\pi}\int_{\delta_{min}}^{\delta_{max}} d \delta~
    \sqrt{1-\left({Y(\delta)-F\over u}\right)^2}\;,\nonumber\\
& & \hspace{0.9in} u\leq u_{sv}(F)  \;.
\label{r(u,F)_1}
\end{eqnarray}

\begin{figure} 
\begin{center}
\epsfxsize=8cm
\epsfbox{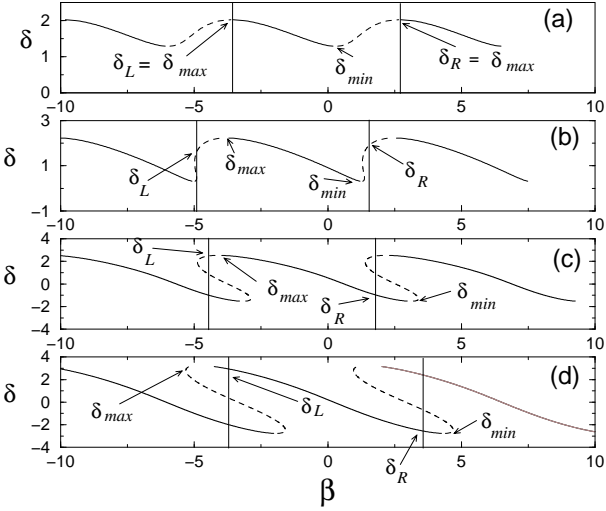}  
\end{center}
\caption{The phase $\delta$ as a function of $\beta$ for various values of $u$
and $F<F_\uparrow^c(u)$. Also shown in each frame are the values 
of $\delta_{min}$ and $\delta_{max}$ defined in Eq.~(\ref{deltamin_deltamax})
and the boundary points $\delta_R$ and $\delta_L$ of the connected portion of
the function $\delta(\beta)$ that is used to evaluate the integrals 
determining the coherence in each case. The four curves corresponds to 
the four cases discussed in the text: (a) $u<u_{sv}(F)$, 
where $\delta(\beta)$ is single-valued.  
In this case we can choose $\delta_L=\delta_{max}$, 
which requires $\delta_R=\delta_{max}$. As $F$ is increased, $\delta_{max}$
grows until $\delta_{max}=\delta_L=\pi$ at $F=F_\uparrow^c$.
(b) $u_{sv}(F)<u\leq u_1(F)$;
(c) $u_{1}(F)<u\leq u_2(F)$; and
(d) $u>u_2(F)$.}
\label{delta(beta,u,F)_fourplots}  
\end{figure} 

For $u>u_{sv}(F)$, the function $\delta(\beta)$ is multivalued. In
this case there are multiple possible metastable states
$\delta(\beta)$. We can use any one of these to calculate $r(u,F)$ as
long as the chosen state satisfies the imaginary part of the
self-consistency condition, and lies in the range $[-\pi,\pi]$, but as 
explained above we choose to focus on the one corresponding to a connected 
portion of $\delta(\beta)$. 
As $u$ is increased at fixed $F$, $\delta_L(u,F)$ increases and
$\delta_R(u,F)$ decreases. For hard pinning forces $\delta_R$ reaches 
$\delta_{min}$ before $\delta_L$ reaches $\delta_{max}$. It is then 
convenient to distinguish three regions.
\begin{enumerate}
\item $u_{sv}(F)<u\leq u_1(F)$, where $u_1(F)$ is the value of $u$ where 
$\delta_R=\delta_{min}$. 
In this region the connected portion includes all of the $\beta_0(\delta)$ 
piece and some of both the $\beta_{-1}(\delta)$ and $\beta_{1}(\delta)$ 
pieces 
as shown in Fig.~\ref{delta(beta,u,F)_fourplots}(b).
The imaginary part of the self consistency condition is then given by
\begin{eqnarray}
0&=&\int_{\delta_L}^{\delta_{max}} d \delta\left({d \beta_{-1} \over d \delta}\right) \sin[\delta+\beta_{-1}(\delta)]\nonumber\\
&+&\int_{\delta_{max}}^{\delta_{min}} d \delta\left({d \beta_0 \over d \delta}\right) \sin[\delta+\beta_0(\delta)]\nonumber\\
&+&\int_{\delta_{min}}^{\delta_R} d \delta\left({d \beta_1 \over d \delta}\right) \sin[\delta+\beta_1(\delta)] \;,
\label{Fscc1}
\end{eqnarray}
with the additional requirement
\begin{equation} 
\beta_{-1}(\delta_L)+2\pi=\beta_1(\delta_R) \;.
\label{deltaLR1} 
\end{equation} 
Once the values of $\delta_L(u,F)$ and $\delta_R(u,F)$ 
have been obtained by solving Eqs.~(\ref{Fscc1}) and (\ref{deltaLR1}),
the function $f(u,F)$, is computed using Eq.~(\ref{Real_appB}),
which now has the explicit form
\begin{eqnarray}
f(u,F)&=&\int_{\delta_L}^{\delta_{max}} d \delta\left({d \beta_{-1} \over d \delta}\right) \cos[\delta+\beta_{-1}(\delta)]\nonumber\\
&+&\int_{\delta_{max}}^{\delta_{min}} d \delta\left({d \beta_0 \over d \delta}\right) \cos[\delta+\beta_0(\delta)]\nonumber\\
&+&\int_{\delta_{min}}^{\delta_R} d \delta\left({d \beta_1 \over d \delta}\right) \cos[\delta+\beta_1(\delta)] \;.
\end{eqnarray}
\item $u_1(F)<u\leq u_2(F)$, where $u_2(F)$ is the value of $u$ where 
$\delta_L=\delta_{max}$. In this region the connected portion includes 
only parts of the $\beta_0(\delta)$ and $\beta_{-1}(\delta)$ pieces 

In this region $\delta_L>\delta_{max}$,
but $\delta_R<\delta_{min}$, as shown in Fig.~\ref{delta(beta,u,F)_fourplots}(c).
The imaginary part of the self consistency condition is then given by
\begin{eqnarray}
0&=&\int_{\delta_{L}}^{\delta_{max}} d \delta\left({d \beta_{-1} \over d \delta}\right) \sin[\delta+\beta_{-1}(\delta)]\nonumber\\
&+&\int_{\delta_{max}}^{\delta_R} d \delta\left({d \beta_0 \over d \delta}\right) \sin[\delta+\beta_0(\delta)]\;,
\end{eqnarray}
where
\begin{equation} 
\beta_{-1}(\delta_L)+2\pi=\beta_0(\delta_R)\;.
\label{deltaLR2} 
\end{equation} 
This pair of equations 
yields $\delta_L$ and $\delta_R$, which can then be used to calculate 
$f(u,F)$ as 
\begin{eqnarray}
f(u,F)&=&\int_{\delta_{L}}^{\delta_{max}} d \delta\left({d \beta_{-1} \over d \delta}\right) \cos[\delta+\beta_{-1}(\delta)]\nonumber\\
&+&\int_{\delta_{max}}^{\delta_R} d \delta\left({d \beta_0 \over d \delta}\right) \cos[\delta+\beta_0(\delta)]\;.
\end{eqnarray}
\item $u>u_2(F)$. In this region the simply connected portion of 
the $\delta(\beta)$ curve only contains part of the $n=0$
branch, and {\em none} of the $n=\pm1$ branches as shown in 
Fig.~\ref{delta(beta,u,F)_fourplots}(d). 
The imaginary part of the self consistency condition reads
\begin{eqnarray}
0&=&\int_{\delta_{L}}^{\delta_{R}} d \delta\left({d \beta_0 \over d \delta}\right) \sin[\delta+\beta_0(\delta)]\;,
\label{Fscc4}
\end{eqnarray}
with 
\begin{equation} 
\beta_{0}(\delta_L)+2\pi=\beta_0(\delta_R)\;,
\label{deltaLR4} 
\end{equation} 
and the function $f(u,F)$ is given by
\begin{eqnarray}
f(u,F)&=&\int_{\delta_{L}}^{\delta_{R}} d \delta\left({d \beta_0 \over d \delta}\right) \cos[\delta+\beta_0(\delta)]\;.
\label{Fscc4_f(u,F)}
\end{eqnarray}
\end{enumerate}

As discussed earlier, the depinning force is defined by 
Eq.~(\ref{Fup_defined}), i.e. it is given by the value 
of $F$ where $\delta_u=\pi$. For all values of $u<u_2$, 
we can obtain a simple analytical expression for $F_\uparrow^c$ 
since in this region 
$\delta_u(u,F_\uparrow^c)=\delta_{max}(u,F_\uparrow^c)=\pi$.
Substituting in Eq.~(\ref{deltamin_deltamax}), we obtain
\begin{equation}
\delta_{max}(u,F_\uparrow^c)=-Y^{-1}(F_\uparrow^c+u)=\pi\;,
\end{equation}
which is easily solved to give
\begin{equation}
F_\uparrow^c(u)=1-u,\hspace{0.2in} u\leq u_2(F_\uparrow^c)\;.
\label{Fup_simpleanswer}
\end{equation}
For $u>u_2$, $\delta_{max}$ is outside the connected portion
of the curve included in the integration and $\delta_u=\delta_L$.
So threshold is reached when $\delta_L=\pi$.
In this case it is convenient to directly solve for 
the depinning threshold by setting $\delta_L=\pi$ and $F=F_{\uparrow}^c$
in the self-consistency condition, which is given by
\begin{equation}
0=\int_{\delta_{L}=\pi}^{\delta_{R}(u,F_\uparrow^c)} d \delta
  \left({d \beta_0(\delta,u,F_\uparrow^c) \over d \delta}\right) 
   \sin[\delta+\beta_0(\delta,u,F_\uparrow^c)]\;,
\label{Fscc4_Fup}
\end{equation}
with
\begin{equation} 
\beta_{0}(\delta_L=\pi)+2\pi=\beta_0\big(\delta_R(u,F_\uparrow^c)\big)\;.
\label{deltaLR4_Fup} 
\end{equation} 
Together these two equations yield $F_{\uparrow}^c(u)$.
In Fig.(\ref{F_T(u)}) we plot $F_\uparrow^c(u)$ vs $u$ 
 for the hard pinning potential, $Y(x)=-(x+x^3)/(\pi+\pi^3)$.

\begin{figure} 
\begin{center}
\epsfxsize=8cm
\epsfbox{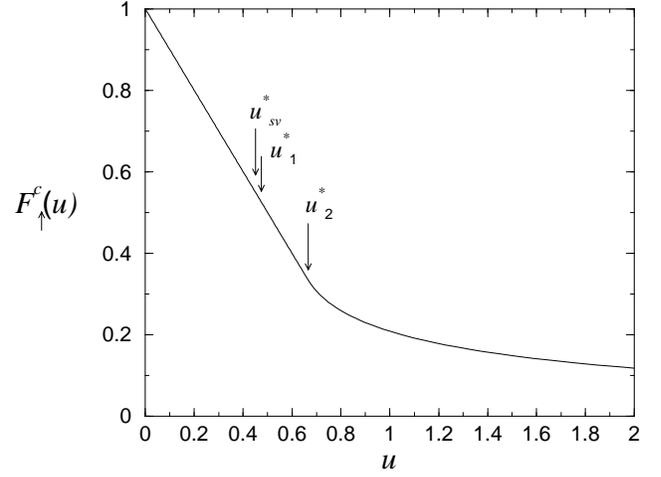}  
\end{center}
\caption{The  force
$F_\uparrow^c(u)$, as a function of $u$ for the hard pinning 
force, $Y(x)=-(x+x^3)/(\pi+\pi^3)$. 
The arrows indicate the values $u_{sv}^*$, $ u_1^*$, and 
$ u_2^*$ separating the four different regions discussed in the text.
These values are defined by the relations 
$u_{sv}(F_\uparrow^c(u_{sv}^*))=u_{sv}^*$, 
$u_1(F_\uparrow^c(u_1^*))=u_1^*$ and $u_2(F_\uparrow^c(u_2^*))=u_2^*$. 
The plot becomes nonlinear beyond $u_2^*$ where the threshold goes from 
being determined by $\delta_{max}=\pi$ to being determined by 
 $\delta_L=\pi$.}
\label{F_T(u)}  
\end{figure}

The method for obtaining $F_{\uparrow}^c(\mu)$ for monotonic soft pinning forces is analogous to that for the hard pinning force, except for one difference. In the case of a monotonic soft pinning force, the value of $\delta_L$ reaches $\delta_{max}$ before $\delta_R$ reached $\delta_{min}$ (the reverse takes place for monotonic hard pinning forces). This means that $u_2<u_1$ so that region 1 is now defined by $u_{sv}<u<u_2$, region 2 by  $u_2<u<u_1$ and region 3 by $u>u_1$. Of course the single valued region remains $u<u_{sv}$. It is not difficult to see that only the expressions for region 2 will differ. In this region the imaginary part of the self consistency condition becomes 
\begin{eqnarray}
0&=&\int_{\delta_{L}}^{\delta_{min}} d \delta\left({d \beta_{0} \over d \delta}\right) \sin(\delta+\beta_{0}(\delta))\nonumber\\
&+&\int_{\delta_{min}}^{\delta_R} d \delta\left({d \beta_1 \over d \delta}\right) \sin(\delta+\beta_1(\delta)) \;,
\label{Fscc3}
\end{eqnarray}
which along with
\begin{equation} 
\beta_{0}(\delta_L)+2\pi=\beta_1(\delta_R)
\label{deltaLR3} 
\end{equation} 
determines $\delta_L$ and $\delta_R$. The expression for $f(u,F)$ in region 2 is now
\begin{eqnarray}
f(u,F)&=&\int_{\delta_{L}}^{\delta_{min}} d \delta\left({d \beta_{0} \over d \delta}\right) \cos[\delta+\beta_{0}(\delta)]\nonumber\\
&+&\int_{\delta_{min}}^{\delta_R} d \delta\left({d \beta_1 \over d \delta}\right) \cos[\delta+\beta_1(\delta)]\;.
\end{eqnarray}

\subsection{$F_\uparrow^c$ For Non-Monotonic $Y(\delta)$}

The method for obtaining $F_\uparrow^c(\mu)$ for non-monotonic
$Y(\delta)$ is analogous to that outlined for monotonic
$F_\uparrow^c(u)$. Matters are complicated, however, by the existence of 
additional unstable solutions of the kind discussed for $F=0$ in Appendix A.
In principle there is no difference in obtaining $F_\uparrow^c(\mu)$; one 
must simply be careful to ensure that only stable solutions are being 
considered. The differences in the calculation are quite technical 
and we spare the reader the details.

\section{Numerics}
\label{Numerics}

To explore the phase diagrams of the mean-field model, we numerically
integrated the equations of motion to determine $v$ and $r$ as a
function of $F$ and $\mu$. As seen in the main text and earlier
appendices, the macroscopic behavior can depend on the preparation of
the initial state. For $N$ degrees of freedom $i=1,2,\ldots,N$, the
$\beta_i$ for most studies were set uniformly, $\beta_i =
\frac{2\pi}{N}i$.  We studied several different initial
conditions. One of the most frequently used was to set all
$\theta_i=\beta_i$ at $F=0$, which prepares the system in the
incoherent static (IS) state, whenever it is stable.  In order to
prepare the system in a static coherent state, all phases would be set
equal to zero. Coherent moving or static states were also prepared by
starting from a high field $F$ with, say, random initial positions
$\theta_i$. (Incoherent moving states were prepared in some portion of
the phase diagram.  When preparing incoherent sliding states, we used
$M^2=N$ degrees of freedom, with $M$ distinct values for $\beta$; the
values of $\theta_i$ for each $\beta$ value were equally spaced in
{\em time} based on the periodic single-particle ($r=0$) solution to
the equations of motion for the given $\beta$.)  Given the initial
conditions, we typically computed $v(F)$ and $r(F)$ at fixed
$\mu$. This was done by integrating the equations of motion
Eq.~(\ref{eom}) using the fourth-order Runge-Kutta scheme. The force
was raised in small discrete steps: after some amount of time $t_{\rm eq}$
at fixed force, $v$ and $r$ are measured and then $F$ is increased
(decreased) some small amount $\delta F$.  With this algorithm, the
time average of the ramp rate $dF/dt$ is given by $\delta F/
t_{\rm eq}$. In some cases, we fixed $F$ and ramped $\mu$ up and down in a
similar fashion.

While the ramp rate and system size does affect the depinning 
force, the force at which $v$ goes from zero to non-zero, 
we find generically that for ramp rates smaller than $10^{-5}$ and 
sizes $N$ greater than $256$, we obtain results for 
both the incoherent and coherent depinning line that are relatively
independent of actual ramp rate or system size and agree with 
analytical calculations.  There is agreement even though the coherent depinning
curve is analytically obtained using the assumption that $u$ is adiabatically 
increased.  For the simulations, on the other hand, $F$ or $\mu$ is 
increased (decreased) slowly. Adiabatically ramping $\mu$ is not necessarily
equivalent 
to adiabatically ramping $u$ since the former does not insure that $r$ 
changes slowly, but we do find the correct coherent 
depinning line by sitting at a fixed $F$ and ramping up $\mu$.  

The analytical analysis in Secs.\ V and VI provides us with the
depinning line as approached from the pinned phase, but it does not
give us insight into the nature of the depinning transition.  For
example, there could be hysteresis in $v(F)$ or $r(F)$ for cyclical
histories in the force, for sufficiently large system sizes and
arbitrarily small ramp rates.  Hysteresis in the order parameters
implies that the depinning transition is discontinuous.  If there is
hysteresis in $v(F)$, then the depinning line as approached from the
moving phase must be different from the depinning line computed in
Secs. VI and V.  To numerically search for hysteresis, we prepare the
system in a coherent moving state and lower the force until the system
stops.  If this repinning line is different from the analytical
depinning result, hysteresis between the static (IS or CS) and moving
phases is present and there is a region where the two phases coexist.

For every potential investigated, we find that there is a range of
$0<\mu<\mu_e$ where there is a coexistence of the moving and stationary
solutions.  In general, there is hysteresis between coherent moving
(CM) and incoherent static (IS) phases.  For the piecewise linear
pinning force, the hysteresis extends into the coherent pinned CS
region.  In other words, the coherence $r$ jumps from one finite value
to another at the depinning transition and there is hysteresis in both
$r$ and $v$ ($F_{\downarrow}\neq F^c_{\uparrow}$). The numerical
evidence for this is shown in Fig. \ref{looparealin}, which shows the
area of the hysteresis loop, $\int_0^\infty
dF\,\left[v^\downarrow(F)-v^\uparrow(F)\right]$, where
$v^\downarrow(F)$ and $v^\uparrow(F)$ are the histories of $v(F)$ for
ramping the field down or up, respectively. The amount of hysteresis,
as measured by this quantity, is independent of system size and
$dF/dt$, which suggests that the simulations are close the adiabatic
and infinite-volume limit. There is a jump down in the area of the
hysteresis loop when $\mu$ exceeds
$\mu_u$, but the area is still non-zero for $\mu >
\mu_u$.

\begin{figure}
\begin{center}
\epsfxsize=8cm
\epsfbox{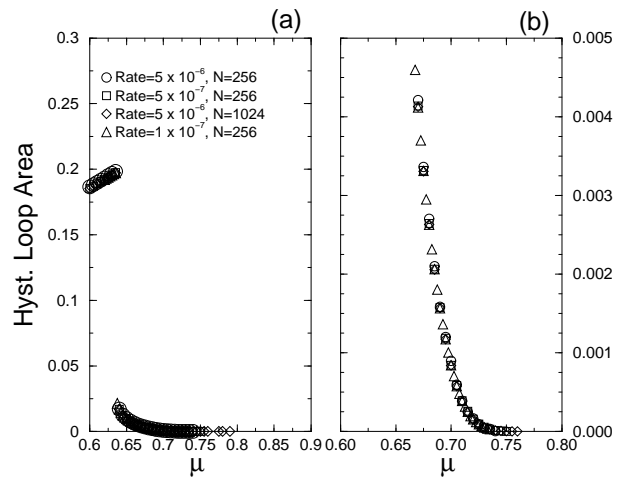}
\end{center}
\caption{\label{looparealin}
(a) Area of the hysteresis loop in $v(F)$ as F is cycled from large
values to zero and then back up again near $\mu_e\approx 0.75$ for the
scalloped potential.  Different system sizes and ramp rates $dF/dt$
are shown. Plot (b) is just a blowup of (a) very near $\mu_e$. }
\end{figure}

For the hard potential, with the history described above,
 $r$ jumps to zero when the system becomes pinned. When the
drive is increased back up again, the system depins at a different
$F_{\uparrow}^i$ when $\mu<\mu_e$.  However, we do not observe
hysteresis between the CM and CS phases.  In fact, the hysteresis
{\em when ramping $F$
vanishes suddenly} at $\mu=\mu_e$.  See Fig. \ref{loopareahard}.  This
is because the slope of the coherent depinning line starts to increase
rapidly at $(\mu_e,F_e)$ and eventually becomes infinite before
curling over to possible
hysteresis.  Above the point at which the slope becomes infinite, the
analytic calculations suggest that
coherent depinning can be observed by increasing $\mu$ at fixed $F$.
This was verified numerically.  For the soft-potential cases
tested we did not observe hysteresis between the CS and CM phases.
Hysteresis is only observed between the IS and CM phases.

\begin{figure}
\begin{center}
\epsfxsize=8cm
\epsfbox{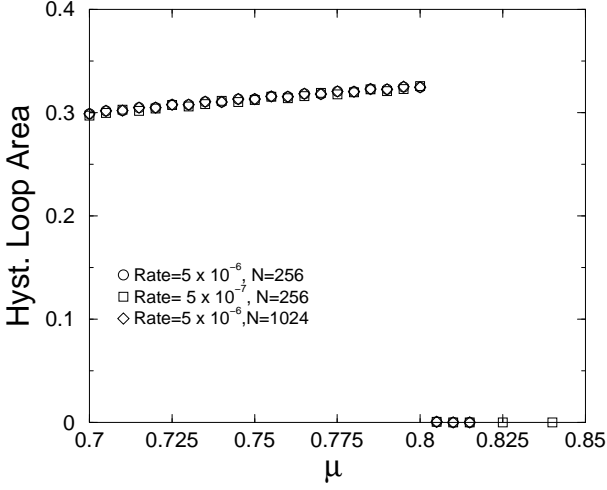}
\end{center}
\caption{\label{loopareahard}Area of the hysteresis loop in $v(F)$ near $\mu_e$ for 
the hard case, where $\mu_e$ is 
the intersection of $F_{\uparrow}^i$ and $F_{\uparrow}^c$.  Different system sizes and ramp rates are shown.}
\end{figure}

\section{Stability of the IM Phase}
\label{IMPhase}

In this section we investigate the existence of a stable incoherent
sliding (IM) phase.  We note that the velocity of a single degree of
freedom is always a periodic function of time.  To obtain a {\em
constant} steady state velocity for a collection of incoherent degrees
of freedom, we assume that at some initial time $t_i$ the $i$-th
degree of freedom is at the minimum of its own potential well, which
in turn is randomly shifted by $\beta$, and perform an average over
the random starting times $t_i$.  These are chosen to be random
variables uniformly distributed on the interval $[0,P]$, with $P$
being the period, that is the time over which the phases advance by
$2\pi$. This procedure guarantees that we sample uniformly all
possible incoherent moving states such that the system reaches a
steady state.

Proceeding as in the study of the incoherent static state, we assume
that the IM phase exists and study its stability.  The
self-consistency condition that must be satisfied by the mean field
solution is given by
\begin{equation}
\label{IM_self}
r(t)e^{i\psi(t)}=\int_P\frac{dt_i}{P}\int^\pi_{-\pi} \frac{d\beta}{2\pi}e^{i\theta(t-t_i,\beta)}\;,
\end{equation}
where the phase $\theta(t-t_i,\beta)$ is the phase at a time $t>t_i$ 
obtained by solving the equation of motion.  In eq.~(\ref{IM_self}), 
both the coherence $r$
and the mean phase $\psi$  are functions of time.  
As in the static stability analysis, we let $\delta(t-t_i,\beta)=\theta(t-t_i,\beta)-\beta$.
We then perturb the IM state 
at the time $t=0$, with a perturbation of the form
\begin{equation}
\delta_p(-t_i,\beta)=\delta_0(-t_i)-\epsilon\sin\left(\beta+\delta_0(-t_i)\right)\;,
\end{equation}
with $\epsilon<<1$.  After inserting this in Eq.~(\ref{IM_self}), we evaluate the right hand side
at $t=0$
to $\mathcal{O}(\epsilon)$, obtaining
$r(t=0)=\epsilon/2$ and $\psi(t=0)=0$.   We then use this to 
compute $\dot{r}(0)$
to linear order, with the result
\begin{equation}
\dot{r}(0)=\frac{\epsilon}{2}\bigg[\frac{\mu}{2}+\int_P \frac{dt_i}{P} Y'(\delta_0
(-t_i))\bigg]\;.
\end{equation}
If $r(t)$ is monotonic in time, then its stability is entirely determined by the sign of
$\dot{r}(0)$. We would then conclude that there is a critical value $\mu_c(F)$ 
of the coupling strength below which the IM phase is stable, with
\begin{equation}
\label{IM_mucrit}
\mu_c(F)=-\frac{2}{P}\int_P \frac{dt_i}{P} Y'(\delta_0
(-t_i))\;.
\end{equation}
With a change of variable from $t_i$ to $\delta_0$ (using the equation 
of motion), one finds that $\mu_c=0$ for all $F>F_{\rm sp}$ for any continuous pinning force 
(since $Y(\pi)=0$).  For discontinuous 
pinning forces, however, we can evaluate the integral in Eq.~(\ref{IM_mucrit}) by
splitting the integral in a contribution from 
the smoothly varying part of $Y'(\delta_0)$ on the interval $[-\pi,\pi)$ and a jump 
at $\delta_0=\pi$. This gives 
\begin{equation}
\mu_c(F) =-\frac{2}{P}\bigg[\ln(\frac{F+|Y(\pi)|}{F-|Y(\pi)|}) -2 \frac{Y(\pi)}{F+|Y(-\pi)|}\bigg]\;,
\end{equation}
where $P$ is a function of $F$. For the piecewise linear force, one 
can evaluate $P$ and find $\mu_c(F)>0$ for some $F>F_{\rm sp}$.  The critical 
value of $\mu_c(F)$ is given by
\begin{equation}
\mu_c(F)=\mu_u\bigg[1-\frac{2}{\ln(\frac{F+a\pi}{F-a\pi})(F+a\pi)}\bigg].
\end{equation}
In the limit of large $F$, $\mu_c$ approaches zero.  As $F$ approaches
$F_{\rm sp}$, on the other hand, $\mu_c=\mu_u$.  In other words, the
IM stability curve abruptly ends at $(\mu_u,F_{\rm sp})$ as there can
be no IM phase for any $F$ less than $F_{\rm sp}$.  A transition from
an incoherent to a coherent moving phase was indeed obtained
theoretically by Vinokur and Nattermann \cite{NV97} in a model of for
layered charge density waves and also observed by Olson et
al. \cite{Olson01} in numerical simulations of layered
superconductors.  For strong disorder, these authors found a
transition as the drive is increased from a $2D$ state of decoupled
moving layers to a $3D$ state where the moving layers become coupled.
Our short time results suggest that a similar transition may occur in
the isotropic system studied here. However, our numerical studies
indicate that this transition may be an artifact of the short time
analysis.  When testing the stability of a system prepared in the IM
phase numerically, we find that $r(t)$ is generally not a monotonic
function of time. Furthermore, a perturbation of strength $\epsilon$
always destabilizes the IM phase in the limit of large system size,
unless the strength of the perturbation is made to decrease with
system size.  Finally, we verified that the IM phase remains unstable
if the somewhat artificial average over the starting times $t_i$ is
replaced by an average over a narrow distribution of pinning
strengths.  Given these numerical findings, we conclude that the IM
phase is typically unstable in the isotropic mean field model studied
here, although of course we cannot rule out that the system could be
prepared in such a state by some special initial condition.


\begin{thebibliography} {99}




\bibitem{Fisher98} D. S. Fisher, Phys. Rep. {\bf 301}, 113 (1998).

\bibitem{DSF85} D. S. Fisher, Phys. Rev. B {\bf 31}, 1396 (1985).

\bibitem{NF92}
O. Narayan and D. S. Fisher, Phys. Rev. B {\bf 46}, 11520 (1992).

\bibitem{nocross} A. A. Middleton, Phys. Rev. Lett. {\bf 68}, 670 (1992).

\bibitem{Ertas96} D. Ertas and M. Kardar, Phys. Rev. B {\bf 53}, 3520 (1996).

\bibitem{Wiese02} P. Le Doussal, K.J. Wiese, and P. Chauve, Phys. Rev. B {\bf 66}, 174201 (2002).

\bibitem{Nattermann} T. Nattermann, S. Stepanow, L.-H. Tang, and H. Leschhorn, J. Phys. II France, {\bf 2}, 1483 (1992).

\bibitem{LittlewoodCoppersmith86} S. N. Coppersmith and P. B. Littlewood, 
Phys. Rev. Lett. {\bf 57}, 1927 (1986) .

\bibitem{LittlewoodCoppersmith87} S. N. Coppersmith and P. B. Littlewood,
Phys. Rev. B {\bf 36}, 311 (1987).

\bibitem{middleton91}
A. A. Middleton and D. S. Fisher, Phys. Rev. Lett. {\bf 66}, 92 (1991).

\bibitem{middleton93}
A. A. Middleton and D. S. Fisher, Phys. Rev. B {\bf 47}, 3530 (1993).
 
\bibitem{MyersSethna93}
C. R. Myers and J. P. Sethna, Phys. Rev. B {\bf 47}, 11171 (1993).

\bibitem{RossoKrauthSimul}
A. Rosso, A. K. Hartmann, and W. Krauth, Phys. Rev. E {\bf 67}, 021602 (2003).


\bibitem{koshelev94}
A. E. Koshelev and V. M. Vinokur, Phys. Rev. Lett. {\bf 73}, 3580 (1994).

\bibitem{giamarchi96}  T. Giamarchi and P. Le Doussal, Phys. Rev. Lett. {\bf 76}, 3408 (1996).

\bibitem{bmr98} L. Balents, M.C. Marchetti, and L. Radzihovsky, Phys. Rev. B {\bf 57}, 7705 (1998).

\bibitem{moon96} K. Moon,  R. T. Scalettar, and G.T. Zimanyi, Phys. Rev. Lett. {\bf 77}, 2778 (1996).

\bibitem{pardo98} F. Pardo, F. de la Cruz, P.L. Gammel, E. Bucher, and D.J. Bishop, Nature {\bf 396}, 348, (1998).

\bibitem{coexistnote} ``Coexistence'' here, within mean field theory, refers to
multistability of the solutions to the equations of motion. As is
typical in mean field theory, there is no natural notion of an
interface in these calculations, so it is not possible for both of
these phases to exist simultaneously in the same {\em mean-field}
sample, though the question of an interface is open in finite
dimensions.

\bibitem{balents95}
L. Balents and M. P. A. Fisher, Phys. Rev. Lett. {\bf 75}, 4270 (1995).


\bibitem{Maeda85}
A. Maeda, T. Fukuyama, and S. Tanaka, Solid. State Comm. {\bf 55}, 951 (1985).

\bibitem{Maeda90}
A. Maeda, N. Notomi, and K. Uchinokura, Phys. Rev. B {\bf 42}, 3290 (1990).

\bibitem{thorne03}
R. E. Thorne, K. Cicak, K. O'Neill and S. G. Lemay, J. Phys. IV {\bf 12},
291 (2002); and references therein. 

\bibitem{foot_inertia}
This can also result from inertial effects.

\bibitem{Strogatz}
S.H. Strogatz, C.M. Marcus, R.M. Westervelt, and R.E. Mirollo,
Phys. Rev. Lett. {\bf 61}, 2380 (1988); Physics D {\bf 36}, 23 (1989).

\bibitem{levy92}
J. Levy, M. S. Sherwin, F. F. Abraham, and K. Wiesenfeld,
Phys. Rev. Lett. {\bf 68}, 2968 (1992).

\bibitem{levy94} A. Montakhab, J. M. Carlson and J. Levy, Phys. Rev. B {\bf 50}, 11227 (1994).

\bibitem{NV97} V. M. Vinokur and T. Nattermann, Phys. Rev. Lett. {\bf 79}, 3471 (1997).

\bibitem{MMP00} 
M. C. Marchetti, A. A. Middleton and T. Prellberg, 
Phys. Rev. Lett. {\bf 85}, 1104 (2000).

\bibitem{MCMKD02} 
M. C. Marchetti and K. A. Dahmen, Phys. Rev. B {\bf 66}, 214201 (2002).

\bibitem{SF01} 
J. M. Schwarz and D.S. Fisher, Phys. Rev. Lett. {\bf 87}, 96107 (2001).

\bibitem{SF03} 
J. M. Schwarz and D. S. Fisher, Phys. Rev. E 67, 021603 (2003).

\bibitem{MMSS03}
M. C. Marchetti, A. A. Middleton, K. Saunders, and J. M. Schwarz, 
Phys. Rev. Lett. {\bf 91}, 107002 (2003).

\bibitem{gruner} G. Gr\"uner, Rev. Mod. Phys. {\bf 60}, 1129 (1998).


\bibitem{ThornePhysicsToday}
R. E. Thorne, Physics Today {\bf 49}, 42 (1996).

\bibitem{coppersmith90}
S. N. Coppersmith, Phys. Rev. Lett. {\bf 65}, 1044 (1990);
Phys. Rev. B {\bf 44}, 2887 (1991).

\bibitem{SCAM91}
S. N. Coppersmith and A. J. Millis, Phys. Rev. B {\bf 44}, 7799 (1991).


\bibitem{myers99}
C. R. Myers, in {\it Structured adaptive mesh
refinement (SAMR) grid methods - The IMA Volumes in Mathematics and
Its Applications}, 117 (Springer-Verlag, 1999).

\bibitem{karttunen99} M. Karttunen, M. Haatja, K. R. Elder, and M. 
Grant, Phys. Rev. Lett. {\bf 83}, 3518 (1999).

\bibitem{switch} R. P. Hall, M. F. Hundley, and A. Zettl, Phys. Rev. 
B {\bf 38}, 13002 (1988).

\bibitem{Thorneexpt} 
T. L. Adelman, J. McCarten, M. P. Maher, D. A. DiCarlo, and R. E. Thorne, 
Phys. Rev. B {\bf 47}, 4033 (1993).

\bibitem{broadband} M. P. Maher, T. L. Adelman, J. McCarten, and 
D. A. DiCarlo, Phys. Rev. B {\bf 43}, 9968 (1991).

\bibitem{lemay98}
S. G. Lemay, M. C. de Lind van Wijngaarden, T. L. Adelman, and R. E. Thorne,
Phys. Rev. B {\bf 57}, 12781 (1998).

\bibitem{JensenBrassBerlinskySimul88}
H. J. Jensen, A. Brass, and A. J. Berlinsky, Phys. Rev. Lett. {\bf
60}, 1676 (1988).

\bibitem{JensenBrassBrechetBerlinskySimul88}
H. J. Jensen, A. Brass, Y. Brechet, and A. J. Berlinsky, Phys. Rev. B
{\bf 38}, 9235 (1988).

\bibitem{NoriSimul96} F. Nori, Science {\bf 271}, 1373 (1996).

\bibitem{ReichardtOlsonSimul} C. J. Olson, C. Reichhardt, and F. Nori, 
Phys. Rev. Lett. {\bf 81}, 3757 (1998).

\bibitem{GBD} N. Gronbech-Jensen, A. R. Bishop, and D. Dominguez, Phys. Rev. 
Lett. {\bf 76}, 2985 (1996).

\bibitem{FMM} M. C. Faleski, M. C. Marchetti, and A. A. Middleton, Phys. Rev. 
B {\bf 54}, 12427 (1996).

\bibitem{Marchevsky97}
M. Marchevsky, J. Aarts, P. H. Kes, and M. V. Indebom,
Phys. Rev. Lett. {\bf 78}, 531 (1997).


\bibitem{lorenz} A. Tonomura, Micron {\bf 30}, 479 (1999).

\bibitem{Troyanovski99}
A. M. Troyanovski, J. Aarts, and P. H. Kes, Nature {\bf 399}, 665 (1999).

\bibitem{STM}
A. M. Troyanovski, M. van Hecke, N. Saha, J. Aarts, and P. H. Kes,
Phys. Rev. Lett. {\bf 89}, 147006 (2002).


\bibitem{fingerprint}
S. Bhattacharya and M.J. Higgins, Phys. Rev. Lett. {\bf 70}, 2617
(1993); S. Bhattacharya and M.J. Higgins, Phys. Rev. B {\bf 52}, 64
(1995); M.J. Higgins and S. Bhattacharya, Physica C {\bf 257}, 232
(1996).

\bibitem{Hellerqvist96}
M. C. Hellerqvist, D. Ephron, W. R. White, M. R. Beasly, and
A. Kapitulnik, Phys. Rev. Lett. {\bf 76}, 4022 (1996);
M. C. Hellerqvist and A. Kapitulnik, Phys. Rev. B {\bf 56}, 5521
(1997).

\bibitem{maeda02}
A. Maeda, T. Tsuboi, R. Abiru, Y. Togawa, H. Kitano, K. Iwaya, and
T. Hanaguri, Phys. Rev. B {\bf 65}, 054506 (2002).

\bibitem{paltiel}
Y. Paltiel, E. Zeldov, Y. N. Myasoedov, K. Shtrikman, S. Bhattacharya,
M. J. Higgins, Z. L. Xiao, E. Andrei, P. L. Gammel, and D. J. Bishop,
Nature {\bf 403}, 398 (2000).

\bibitem{Paltiel02}
Y. Paltiel, Y. Myasoedov, E. Zeldov, G. Jung, M.L. Rappaport,
D. E. Feldman, M.J. Higgins, and S. Bhattacharya, Phys. Rev. B {\bf
66}, 60503 (2002).

\bibitem{marchevsky}  M. Marchevsky, M. J. Higgins, and S. Bhattacharya, 
Phys. Rev. Lett. {\bf 88}, 87002 (2002).

\bibitem{Gorkov}
L. P. Gor'kov, Pis'ma Zh. Eksp. Teor. Fiz.
{\bf 38}, 76 (1983) [JETP---Lett. {\bf 38}, 87 (1983)].

\bibitem{OngMaki}
N. P. Ong and K. Maki, Phys. Rev. B {\bf 32}, 6582 (1985).

\bibitem{inui88}
M. Inui, R. P. Hall, S. Doniach, and A. Zettl, Phys. Rev. B {\bf 38}, 
13047 (1988).

\bibitem{AAMthesis} A. A. Middleton, Ph.D. thesis, Princeton University, 1990.


\bibitem{kuramoto}
Y. Kuramoto, {\it Chemical Oscillations, Waves, and Turbulence}
(Springer, Berlin, 1984).


\bibitem{FLR}
H. Fukuyama and P. A. Lee, Phys. Rev. B {\bf 17}, 535 (1978);
P. A. Lee and T. M. Rice, Phys. Rev. B {\bf 19}, 3970 (1979).

\bibitem{sinfootnote}
In the case $Y(x)=\sin(x)$ it is possible to solve for
$\delta(\beta)$. This is demonstrated in \onlinecite{Strogatz}.
 
\bibitem{adiabatic_u_footnote}
The astute reader will question the suitability of choosing to
evaluate $r$ in an adiabatically increasing $u$ scenario. This is
because perturbation theory has already shown that for ``soft'' and
piecewise linear pinning forces it is {\em not} possible to
adiabatically increase $u=r \mu$ due to the existence of a jump in
$r$. Numerical analysis shows however that the metastable states
accessed through adiabatically increasing $\mu$ are the same as those
that would be accessed were it possible to adiabatically increase $u$.

\bibitem{F_up_footnote}
Note that there are two functions $F_\uparrow^c(\mu)$ which represents
the dependence of the coherent depinning threshold on $\mu$ and
$F_\uparrow^c(u)$ which represents the dependence of the coherent
depinning threshold on $u$. These two functions are distinguished by
their arguments. To avoid confusion it may be preferable to denote the
latter as ${\tilde F}_\uparrow^c(u)$, then $F_\uparrow^c(\mu)$ can be
written as ${\tilde F}_\uparrow^c(u=\mu r(\mu))$. However, since this
leads to cumbersome nomenclature, we do not use it here.

\bibitem{Olson01} C. J. Olson, C. Reichhardt and V.M. Vinokur, 
Phys. Rev. B {\bf 64}, 140502 (2001).

\bibitem{Ling}
A. Pertsinidis and X. S. Ling, {\it Bull. Am. Phys. Soc.} {\bf 48} (2001).

\bibitem{ReichhardtOlson}
C. Reichhardt and C. J. Olson, Phys. Rev. Lett. {\bf 89}, 078301
(2002).
\end{thebibliography}
\end{document}